\documentclass[preprint]{aastex}
\usepackage{graphicx}
\shorttitle{Recovering the Nonlinear Biasing Function}

\newcommand{\ltsima}{$\; \buildrel < \over \sim \;$}
\newcommand{\lsim}{\lower.5ex\hbox{\ltsima}}
\newcommand{\gtsima}{$\; \buildrel > \over \sim \;$}
\newcommand{\gsim}{\lower.5ex\hbox{\gtsima}}

\begin{document}

\title{On Recovering the Nonlinear Bias Function from
Counts in Cells Measurements}

\author{Istv\`an Szapudi and Jun Pan}
\affil{Institute for Astronomy, University of Hawaii, Honolulu, HI 96822}

\begin{abstract}
We present a simple and accurate
method to constrain galaxy bias based on the distribution of
counts in cells. The most unique feature of our technique
is that it is applicable to non-linear scales, where both
dark matter statistics and the nature of 
galaxy bias  are fairly complex. First, we estimate
the underlying continuous distribution function from 
precise counts-in-cells measurements assuming local 
Poisson sampling. Then a robust, non-parametric inversion of the 
bias function is recovered from the comparison of the cumulative
distributions in simulated dark matter and galaxy catalogs. 
Obtaining continuous statistics from the discrete
counts is the most delicate novel part of our recipe.
It corresponds to a deconvolution
of a (Poisson) kernel. For this we present two alternatives:
a model independent algorithm based on Richardson-Lucy iteration, 
and a solution using a 
parametric skewed lognormal model. We find that the
latter is an excellent approximation for the dark matter distribution,
but the model independent iterative procedure is more suitable for
galaxies. Tests based on high resolution dark matter simulations
and corresponding mock galaxy catalogs show that we can reconstruct
the non-linear bias function down to highly non-linear scales
with high precision  in the range of $-1 \le \delta \le 5$.
As far as the stochasticity of the bias, we have found a 
remarkably simple and accurate formula based on Poisson
noise, which provides an excellent approximation for the scatter
around the mean non-linear bias function.
In addition we have found that redshift distortions have a 
negligible effect on our bias reconstruction, therefore our
recipe can be safely applied to redshift surveys.

\end{abstract}


\section{Introduction}

The principal aim of statistical analysis 
of galaxy catalogs is to extract information about the initial fluctuations
in the early universe, their subsequent gravitational growth, 
and processes of galaxy formation. To decipher the available data,
the distribution of the underlying dark matter have to be inferred
from the distribution of galaxies. The two distributions in principle
can be assumed to be quite different, i.e.
galaxies are biased tracers of the underlying dark matter
statistics \citep{kaiser84, bbks86, ls92, dl99}. 
In a class of phenomenological models, 
galaxy fluctuations $\delta_g$ are assumed to be
a function of the matter fluctuation field, $f(\delta_m)$,
including the simplest case of linear bias  $\delta_g=b\cdot \delta_m$
\citep{fg93, szapudi95, matsubara95, matsubara99}.
Accurate knowledge of this function is needed to interpret
galaxy statistics in light of theories of structure formation.

The bias function can be extracted from measurements of
large scale structure statistics and Cosmic Microwave
Background (CMB).
One possibility is to use second order statistics 
from several sources, especially CMB maps
together with galaxy catalogs to constrain bias 
(e.g. \citealt{efstathiou02}; \citealt{lahav02}; \citealt{verdewmap03}).
The underlying idea is that fluctuations in the CMB have
a direct relationship with density fluctuations in the
early universe. The comparison with present day galaxy
statistics reveal information on both the gravitational amplification, 
and galaxy formation. The application of these methods is
limited to fairly large scales, since non-linear growth and
bias for galaxies, as well as secondary anisotropies for the CMB
become more and more complex on smaller scales.

Another class of methods use higher order statistics 
\citep{fg93, gf94, fry94, szapudibias98, 3fs01, verde2df02} or 
velocity information (e.g. \citealt{bzpd00}; \citealt{zbhd02})
from galaxy catalogs to derive the bias function internally. 
These methods again are limited to
linear to weakly non-linear scales, where both gravitational
instability and bias theory is on a well understood.
Since the most reliable large scale structure data are
available on smaller scales, it is natural to seek methods
for constraining bias, applicable
on small, non-linear scales.
The principal aim of this work is to introduce such technique
based on a direct comparison of counts in cells in simulations
and data. 

We generalize a simple and elegant idea
by \citet{sbd2000}, based on 
the relation between the (continuous) 
cumulative probability distribution functions of the galaxy and 
matter density fluctuation fields, $C_g(\delta_g)$ and $C_m(\delta_m)$,
\begin{equation} 
C_m(\delta_m)=C_g(f(\delta_m))\ ,
\end{equation}
where $C(\delta)\equiv\int_{-1}^{\delta} p(\delta') d\delta'$.
The above relation allows in principle the recovery of the bias
function, if both cumulative distributions are known:
\begin{equation}
\delta_g=f(\delta_m)=C_g^{-1}\left[ C_m(\delta_m) \right] \ .
\end{equation}
\citet{sbd2000} used an approximate form of this relation by
simply replacing the cumulative probability distribution with
the cumulative distribution of counts in cells 
$C_{N\le N_{max}}=\sum_{N<N_{max}}P_N$, and postulating
that the dark matter cumulative distribution is well described
by a lognormal function. Both of these approximations render
the original form of this method fairly approximate. Dark
matter distribution does not exactly follow a lognormal
distribution (as we show later), and discreteness effects
(the difference between the continuous and discrete distribution)
are important but for the densest galaxy catalogs.
We improve the original idea on both counts: we {\em measure}
the cumulative distribution for both galaxies, and for
dark matter; the latter from simulations with the appropriate
cosmology. In addition, we achieve accurate
reconstruction of the {\em continuous} probability distribution function
from the discrete counts in cells distribution. This eliminates
most of the ``stochastic'' component of the bias arising from 
Poisson noise, and improves the accuracy of the reconstruction.
Moreover, this means that we can penetrate smaller scales,
where discreteness effects are important.

In what follows, we show that  our method can recover
the bias in a robust fashion with unprecedented precision, even in
the non-linear regime. In the next section we present our method
to estimate the continuous cumulative distribution function from
discrete counts in cells measurements, and thus recover the bias 
function. In section three we test our technique in a suit
of $N$-body simulations in both real and redshift spaces. The
final section contains conclusions and discussions of our
results.

\section{Methods}

The technical challenge of the idea outlined in the previous section
consists of estimating the continuous probability distribution
function from counts in cells measurements (CIC) 
in a robust way. Note that there are several proved
and fast methods to estimate CIC distributions from galaxy catalogs or
in simulations 
( \citealt{szapudi98}; \citealt{sqsl99}; \citealt{cs03}). 
In this work we are using the former method for all measurements.

The count probability distribution function (CPDF) $P_N$ is the
probability that a certain cell has $N$ objects (galaxies). It
is directly related to the continuous function under the locally
Poissonian approximation
\begin{equation}
P_N=\int_{-1}^{+\infty} p(\delta) \frac{[\langle N \rangle(1+\delta)]^N 
e^{-\langle N \rangle (1+\delta)}}{N!} d\delta \ ,
\end{equation}
where $\langle N \rangle$ is the mean CIC. Some
models of galaxy formations predict sub-Poissonian scatter
for very small halos (e.g. \citealt{gifbias01}; \citealt{cmsb02}; 
\citealt{bw02}).
As long as a specific theoretical model, written
in a convolution form similar to the above, is available, our
technique can be easily adapted. For most scales, however, 
the local Poissonian assumption appears to be correct,
and we assume it for the rest of this work.

As long as Eq (3) can be inverted, Eq (2) can be used to calculate
the bias. The inversion, however, is a fairly delicate process.
Most of the technical challenge of our aim was met by devising
pair of new methods
for deconvolving this somewhat unstable equation in a robust way.

\subsection{Richardson-Lucy Deconvolution}

To invert Eq (3) in a model independent way, 
we use the Richardson-Lucy (RL) method (see appendix C in \citealt{bm98} 
and reference there in). This iterative method is based on
the Bayes theorem. The kernel in Eq.(3) is Poissonian
\begin{equation}
K(N,\delta)=\frac{[\langle N \rangle (1+\delta)]^N e^{-\langle N \rangle 
(1+\delta)}}{N!} \ .
\end{equation}
Since $N$ is within $[0,N_{max}]$ in practice, the $K(N,\delta)$ has to be
normalized to unity with respect to $N$. In the probabilistic 
spirit of this method, the 
functions need rescaling: $\hat{K}=K/\sum_N K$ and $\hat{p}=p\cdot\sum_N K$.
Starting from an initial guess of $\hat{p}_1$,  we can calculate
$P_{N,1}$ via Eq.(3). From this, a better approximation of 
$\hat{p}$ is obtained,
\begin{equation}
\hat{p}_2=\hat{p}_1 \sum_{N=0}^{N_{max}}\frac{P_N}
{P_{N,1}} \hat{K}(N,\delta) \ .
\end{equation}  
This in turn is used as the input for the next iteration. The improvement
of the fit $P_{N,i}$ after the $i^{th}$ iteration is quantified by
the cost function $\chi^2=\sum_N(P_N/P_{N,i}-1)^2$, where
$P_N$ is the measured CIC distribution.
 One caveat is the phenomenon of
``over-learning'', when the recovered probability distribution
starts to fit small fluctuations in the measured $P_N$. This can
inject artificial features into the results after (too) many iterations.
Numerical experiments indicate, that after about
$10$ iteration the  $\log \chi^2$ is becoming fairly small
and changes little. The process is illustrated in Figure 1., where
a lognormal model for the PDF was used to generate
a CPDF curve with noise generated at $5\%$ level. The inverted PDF on
the figure is fairly accurate after 12 iteration,
while the results of 100 iterations clearly show the sign of over-learning.
Figure 2. displays $\log\chi^2$ as a function of iteration: indeed,
beyond 10-12 iteration it is hardly changing; this is the point
of diminishing returns, after which over-learning kicks in.
The displayed cost function is fairly typical, and we have found that
an inspection of $\log\chi^2$ always allows a simple determination
of a sensible stopping point.  Because of the slow change
in the cost function, the inversion is only logarithmically
sensitive to finding the optimal number of iterations, therefore
this prescription is fairly robust.

\begin{figure}
\plotone{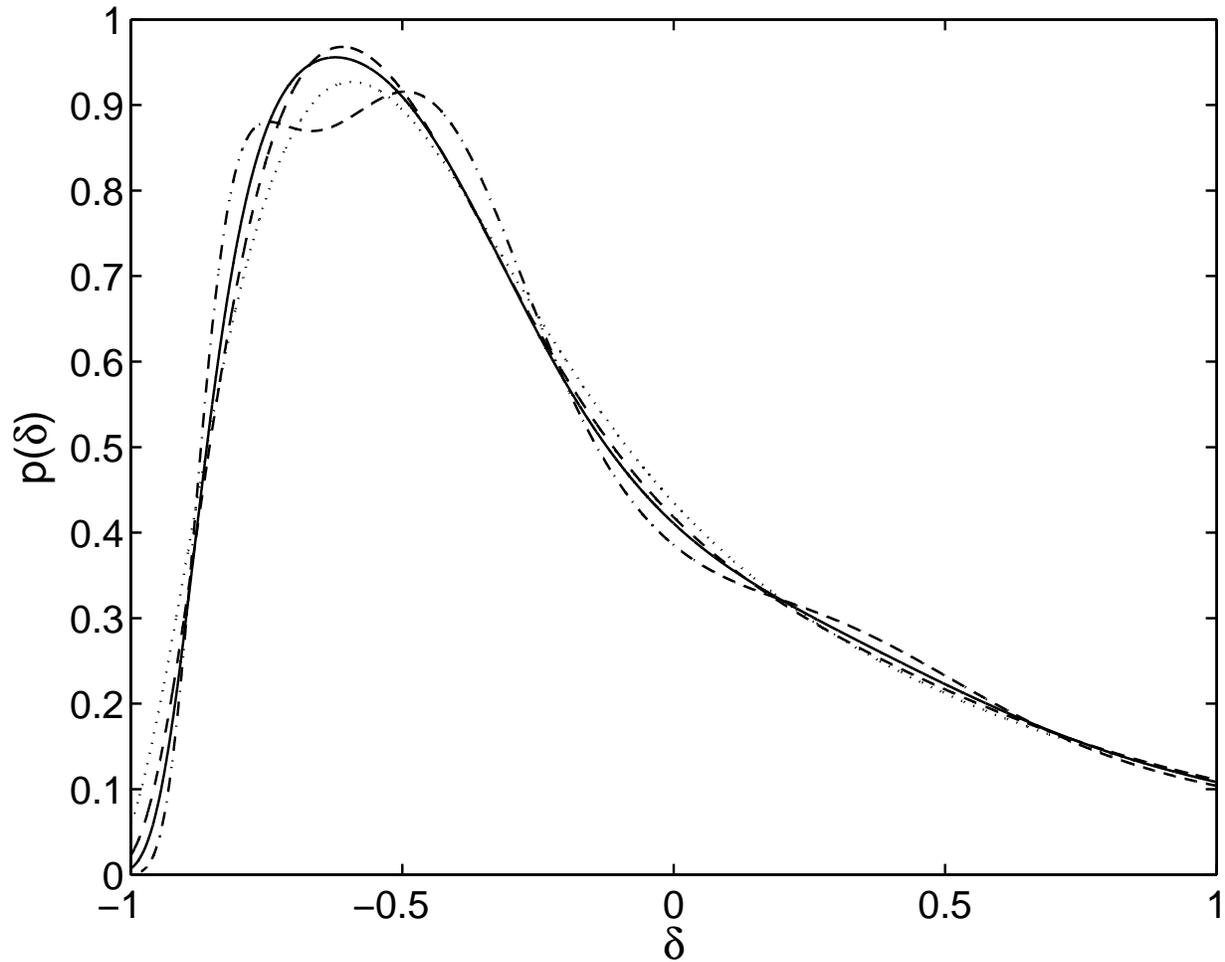}
\caption{Richardson-Lucy inversion of an artificial PDf
is plotted for a number of different iterations.
The solid line corresponds lognormal input model from which the CPDF
was generated according to Eq.(3). Noise was added to the CPDF
at the $5\% $ level.
The dotted line is the result after 5 iterations, the dash line 
corresponds to
12 iterations, while the dash-dots display 100 iterations. After
100 iterations over-learning problems appear (see Figure 2. as well)}
\end{figure}

\begin{figure}
\plotone{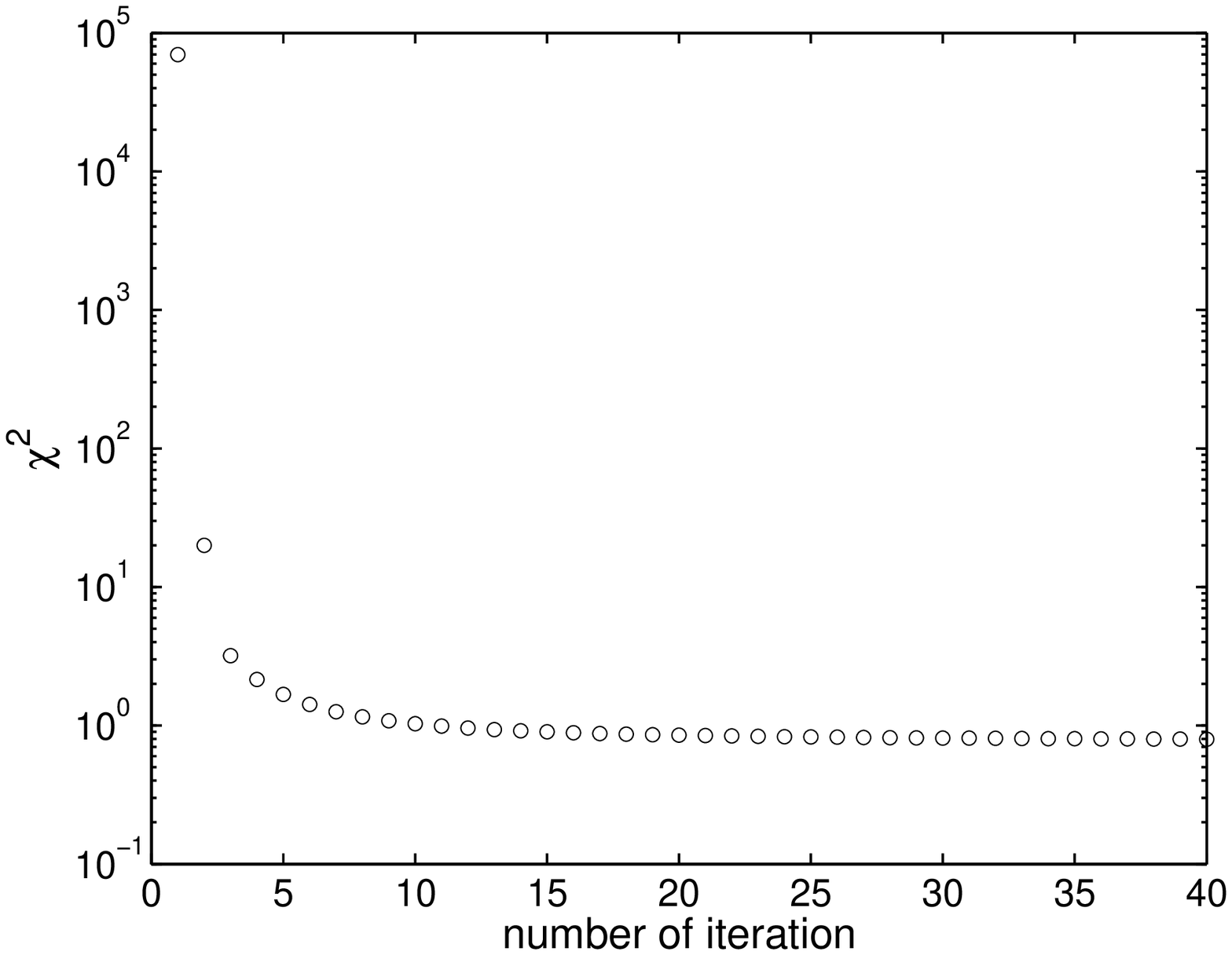}
\caption{The cost function $\log\chi^2$ 
of the Richardson-Lucy method (see text) as a function of
iterations for the example of Figure 1. Note that the change
of the cost function becomes very slow after about 10-15 iterations,
which would be a sensible stopping range for the iterations.}
\end{figure}

Numerical experiments show (see next section) that this method
converges fairly fast and arrives at a robust result if 
$\langle N \rangle \gsim 0.1$. For smaller average counts
convergence slows down such that a few 100 iterations are needed, 
as well as Poisson noise starts to dominate. We recommend that
our method is used down to this value for robust results.

Another difficulty of the RL inversion is purely computational.
Typically $N_{max}\sim 10^4$ for a CPDF of 
dark matter dataset from an N-body simulation, and the kernel is a broad
function. For each pair of $N$ and $P_N$, we need to sample at least
one point at the peak of the kernel $N/\langle N \rangle -1$, and
a few hundred points on each side to keep the integration
accurate. The last step thus costs about $10^{10}$ calculations of the
kernel. Storing the kernel would require
an array of floats of dimension $10^4 \times 10^6$. 
This is unfeasible on a typical small workstation available to us.
Similar computational problems arise with other direct
inversion methods, such as singular value decomposition, etc.
This motivates us to define an alternative, model-dependent method,
which i) requires more modest computational resources, ii) it can
extend toward even smaller scales, i.e. sparser sampling of the
distribution.

\subsection{Skewed Lognormal Model Fit}

An alternative which naturally lends itself is to use
physically motivated parametrized PDF models. This idea was explored by
\citet{ks98},who adopted Gaussian Edgeworth expansions to the third order as
model for $p(\delta)$ to deconvolve Eq.(3) and thereby fit for
skewness and kurtosis. According to their
investigations, such method is severely limited to the weakly
non-linear regime, where the Gaussian Edgeworth expansion is 
good approximation. This is especially important in our application,
where not only the first few moments are interesting, but the
full behavior of the PDF containing information on very high
order moments. Therefore we conclude that the Gaussian Edgeworth
expansion is not a viable model for our purposes.

A more promising empirical model can be built based on
the lognormal distribution (e.g. \citealt{cj91}). This model
is physically motivated, arising naturally from perturbation
theory of the logarithm of the dark matter density field
\citep{sk03}.
Indeed,  the skewed lognormal distribution (SLN3) 
appears to be an excellent approximation for the
PDF of the dark matter field in a wide
range of scales \citep{colombi94,uy96}. It is plausible then,
that SLN3 might be a good approximation
for galaxy distributions as well, as long as the bias is moderate;
the tests of the next section indeed confirm
this idea. Next we outline how one can proceed to 
invert Equation (3) under the assumption of SLN3.

With the notation $\rho=1+\delta$ (the density field), 
$\Phi=\log\rho -\langle \log \rho \rangle$ and $\sigma_\Phi=\langle
\Phi^2 \rangle$, the SLN3 model reads as
\begin{eqnarray}
 p_3(\delta)d\delta& = \Big[ 1+\frac{1}{3!} T_3 \sigma_\Phi H_3(\nu)+
\frac{1}{4!} T_4 \sigma_\Phi^2 H_4(\nu) \nonumber \\&
+ \frac{10}{6!}T_3^2\sigma_\Phi^2 H_6(\nu)\Big] G(\nu)d\nu \ ,
\end{eqnarray} 
where $\nu\equiv \Phi/\sigma_\Phi$,  $H_m(x)$ is an Hermite polynomial
of degree $m$, and $G(x)$ is a Gaussian with zero mean and variance of unity.
The quantities $T_3$ and $T_4$ are the renormalized skewness and 
kurtosis of the field $\Phi$, respectively \citep{colombi94}:
\begin{equation}
T_3=\frac{\langle \Phi^3 \rangle}{\sigma_\Phi^4}\ \ \ ,\ \ \  
T_4=\frac{\langle \Phi^4 \rangle -3\sigma_\Phi^4}{\sigma_\Phi^6} \ .
\end{equation}

The convolution of $p_3(\delta)$ with the 
Poisson kernel, $\widetilde{P}_N$,  corresponds to a model of 
observed $P_N$ depending on four parameters
$\langle \log \rho \rangle$, $\sigma_\Phi$, $T_3$ and $T_4$.
These parameters can be fitted to the observed distribution,
which in turn yields the best approximation 
of the real PDF by $p_3(\delta)$. The following 
Poisson likelihood
function was used to fit the parameters:
\begin{equation}
{\mathcal L}=\prod_N \frac{(MP_N)^{M\widetilde{P}_N} e^{-MP_N}}
{(M\widetilde{P}_N)!} \ ,
\end{equation}
where  $M$ be the total number of cells used for the CIC measurements.
Minimization of $-\ln {\mathcal L}$ with respect to 
$\langle \log \rho \rangle$,
$\sigma_\Phi$, $T_3$ and $T_4$ corresponds to
a third order SLN fit, which is realized with the Powell's method \citep{press92}.

Note that we have experimented with other cost functions for fitting the
parameters, including the conventional minimal $\chi^2$ and a likelihood 
function of \citet{ks98}, similar to the above, but
$MP_N$ replaced with $M\widetilde{P}_N$. Numerical
experience suggests that the above cost function, which
is close to $\chi^2$ for large $MP(N)$, gives the best bias 
reconstruction among the variations tested.

\section{Application to N-body Simulations}

\subsection{Simulations}

The above described bias reconstruction method, including
both inversion methods at its core,  was extensively
tested in a suit of the dark matter 
mock galaxy catalogs extracted from N-body simulations by the GIF project
of the Virgo Consortium \citep{kcdw99}. 
We used the $z=0$ output for a LCDM universe with
$\Omega_m=0.3$, $\Omega_\Lambda=0.7$,
shape parameter, $\Gamma=0.21$, $\sigma_8=0.90$ and $h=0.7$,
force soft length $20h^{-1}$kpc, the simulation
box was of $L=141.3h^{-1}$Mpc. The simulations have
$256^3$ particles of mass $1.4\times 10^{10}M_\sun h^{-1}$.
We used two mock galaxy catalogs, the GIF galaxy catalog and the
GIF galaxy catalog for SDSS. These artificial galaxy catalogs
have been produced using semianalyical galaxy formation models
\citep{kcdw99, somerville00,benson00}. 
They have been studied extensively to explore
biasing as function of luminosity, scale and redshift by \citet{gifbias01}.

Because of the computational difficulty of the RL
inversion for the large dark matter simulations, we used the 
SLN3 model fit exclusively for the inversion of the their PDFs. 
We found that above scales $2.21h^{-1}$Mpc SLN3 provides an
excellent model for the dark matter distribution, while
on smaller scales our model has more and more difficulty to
fit the tail of the distribution (see Figure 3).
On scales of 2.21 $h^{-1}$Mpc, the
measured  CPDF develops a very long tail even in log-space, which
is not well fit by the polynomial correction of the SLN3
beyond $N>\sim 10^3$. However, since $\langle N\rangle=64$ and we aim
to use our method up to $\delta<5$,  we conclude that 
SLN3 will be a good approximation for our purposes.

\begin{figure}
\plotone{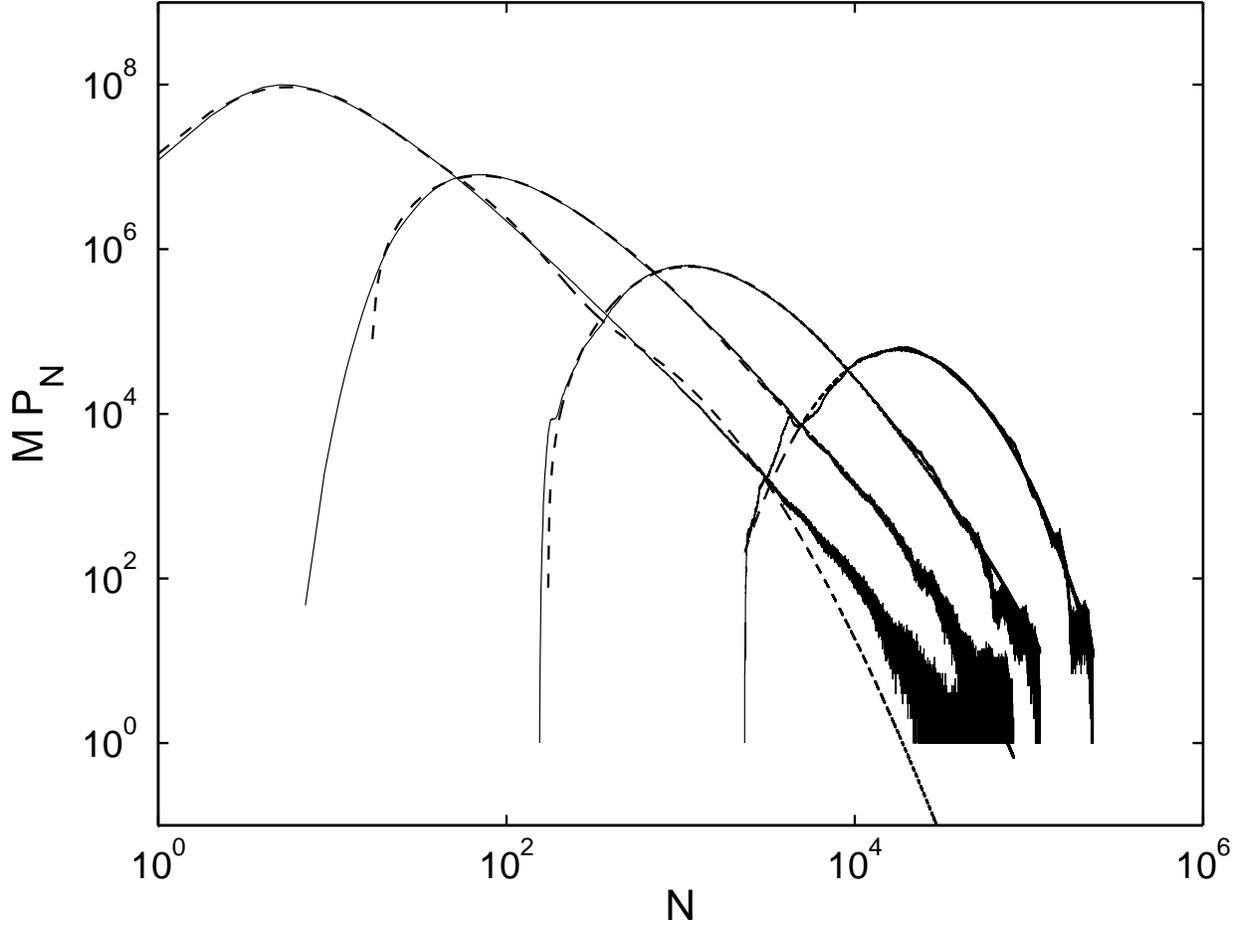}
\caption{The CPDFs of the diluted dark matter sample on four 
different scales, from left to right, 2.21, 4.42, 8.83 and 17.66 
$h^{-1}$Mpc.
Solid lines show the measurements and dash lines represent our
SLN3 model fits.}
\end{figure}

\begin{deluxetable}{crrrrr}
\tablecaption{parameters of SLN3 for the dark matter sample}
\tablewidth{0pt}
\tablehead{
\colhead{Cell Size R ($h^{-1}$Mpc)} &
\colhead{ $\langle N \rangle $ } &
\colhead{ $\langle \log \rho \rangle$ } &
\colhead{ $\sigma_\Phi$ } &
\colhead{ $T_3 \sigma_\Phi$ }&
\colhead{ $T_4 \sigma_\Phi^2$ }
}
\startdata
2.21  & 64 & -1.157 &    1.185 &    1.110  &   1.289 \\
4.42  & 512 &-0.798 &    1.108 &    0.757  &   0.781 \\
8.83  & 4096 &-0.463 &    0.906 &    0.397  &   0.300 \\
17.66  & 32768 & -0.201 &    0.633 &   -0.008  &  -0.080 \\
 \enddata
\end{deluxetable}

\subsection{A null-test of bias reconstruction}

In order to test the reliability of the bias extraction based
on the SLN3  model fit, we selected several subsamples from the 
dark matter simulation by by randomly sampling them at 10\%, 1\% and
0.1\%. The null-test consists of recovering $b=1$ from these catalogs;
our method passed with flying colors.

As a typical example, fits to the measured 
CPDFs at 1\% dilution level are presented
on Figure 4. The fits are excellent approximation to the measurements.
Then the recovered $C(\delta)$ is used to reconstruct the bias of our
diluted subsamples with respect to the full sample. Except for
$R=2.21h^{-1}$Mpc which has a $\sim 10\%$ peak difference, Figure 5.
shows that
the recovered bias is at most  $\sim 5\%$ off from  $b=1$ 
for $\delta<5$.
We have performed the same study with several realizations, 
all giving similar results.
The recovered
bias at $R=2.21, 4.42, 8.83$ and $17.66h^{-1}$Mpc are 
$b=0.96\pm0.12, 1.02\pm0.02, 1.00\pm0.03, $ and $1.01\pm0.02$ respectively. 
\begin{figure}
\plotone{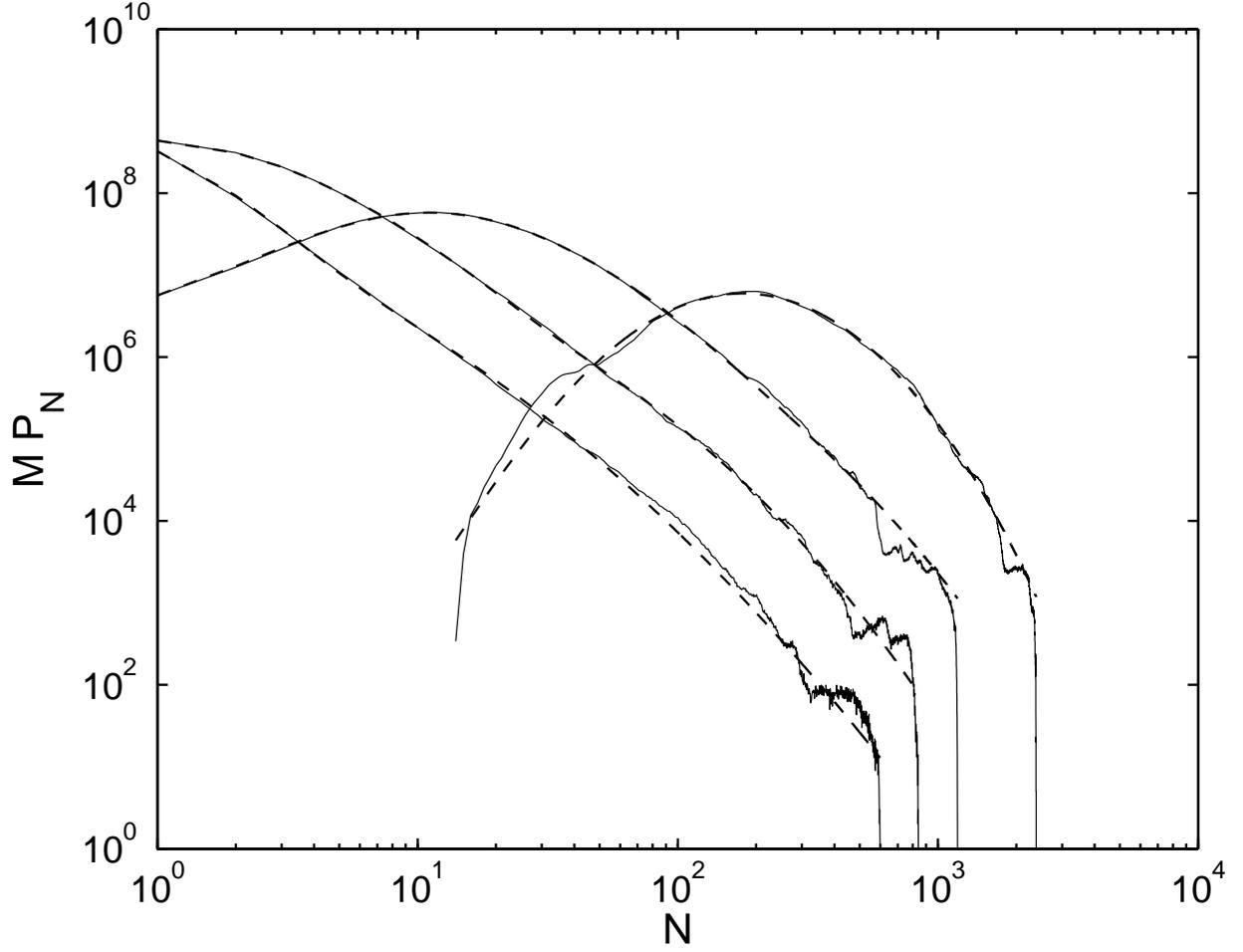}
\caption{The CPDFs of the 1\% diluted dark matter sample on four 
different scales of 2.21, 4.42, 8.83 and 17.66$h^{-1}$Mpc 
(from left to right).
 Solid lines correspond to our measurements,
dash lines represent our model fits.}
\end{figure}

\begin{figure}
\plotone{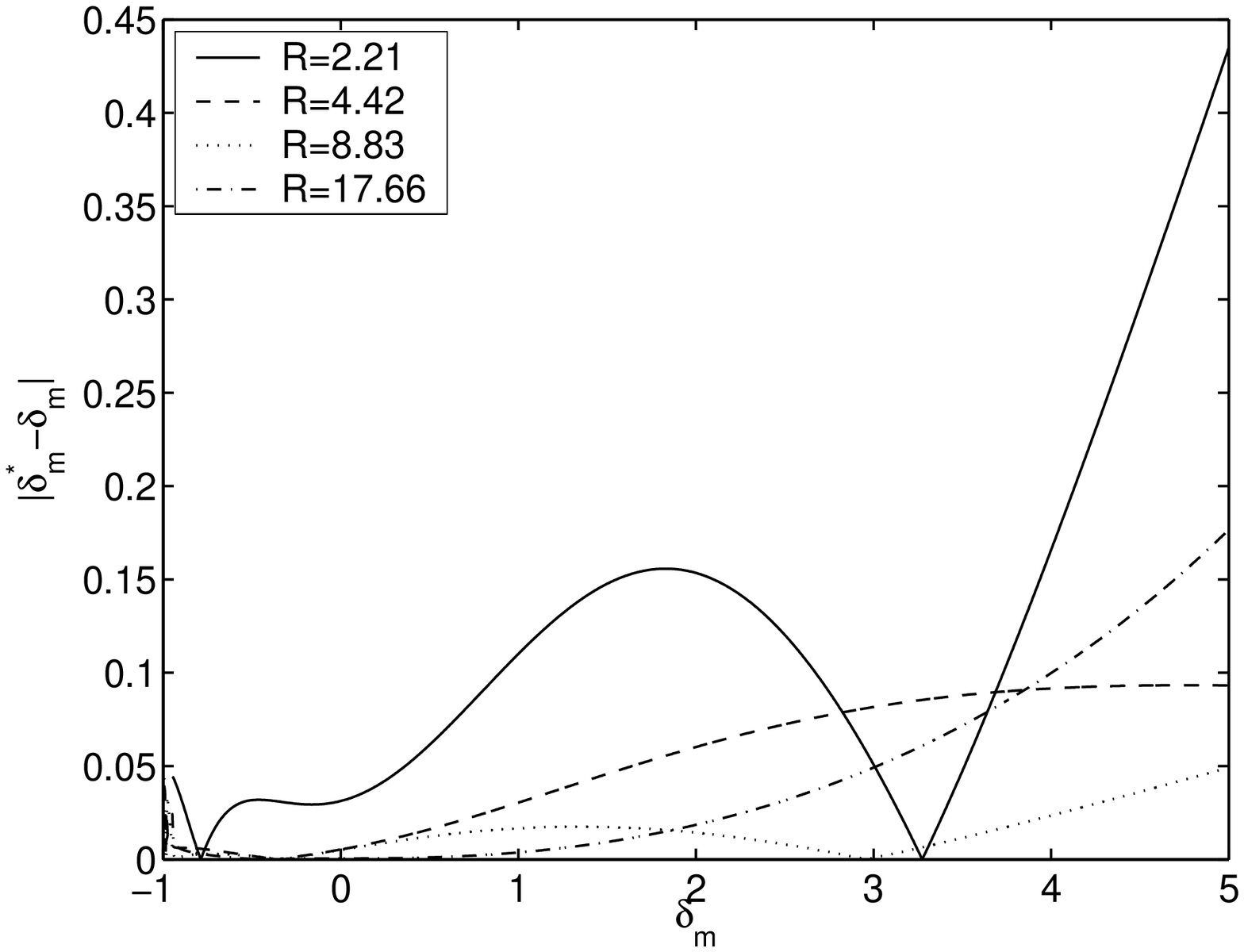}
\caption{The deviation of the recovered bias function 
from $b=1$.$|\delta^*_m-\delta_m|=0$, where $\delta_m^*$ is the
density fluctuation of the 1\% diluted subsample, 
as a function of the density fluctuation $\delta_m$ 
of the full dark matter sample. Error-free reconstruction would
yield 0. Both PDFs are generated by SLN3 model fits.}
\end{figure}

In addition, we have applied the 
Richardson-Lucy method to the 1\% diluted subsamples. 
For $\delta < \sim 4$, the inverted PDFs 
are identical to the PDFs by SLN3 model fits except 
for the scale of $2.21h^{-1}$Mpc. Apparent differences
occur at large $\delta$. When the SLN3 model PDF is used
as a reference of the full dark matter sample, the bias parameter 
recovered by Richardson-Lucy inversion are
1.16, 1.04, 1.03, 0.94 on scales of $R=2.21, 4.42, 8.83, 17.66h^{-1}$Mpc,
respectively for $\delta <5$. From Figure 6. it is clear that on large scales
both methods are in good agreement. On smaller scales the deviation
is due to the domination of discreteness effects.

\begin{figure}
\plotone{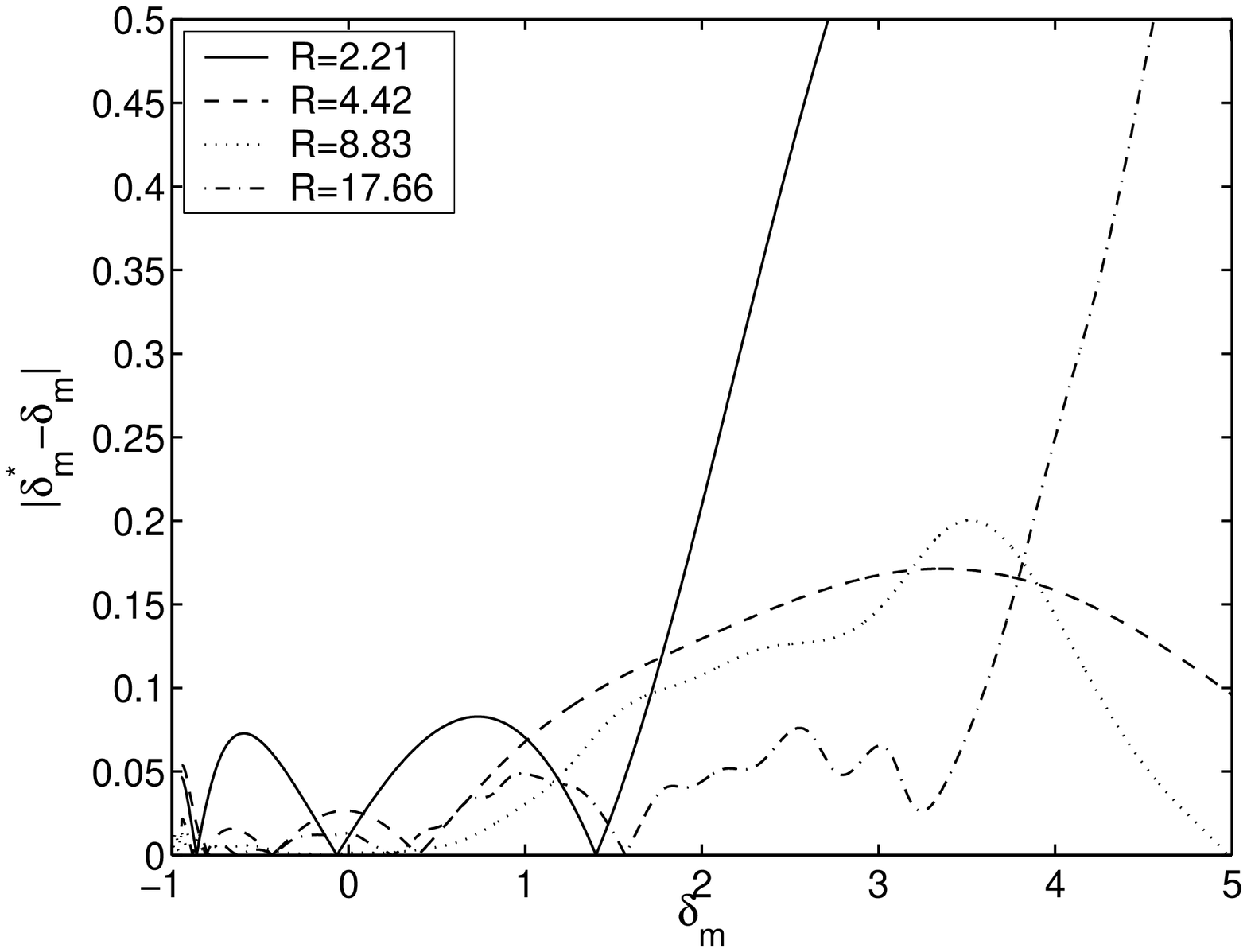}
\caption{ Same as Figure 5, except the PDFs of
diluted subsample are inverted by Richardson-Lucy method, while
the PDFs of the full dark matter simulation are still modeled by SLN3}
\end{figure}

\subsection{Bias from the GIF Mock Galaxy Catalog}

In this section we subject our method to extensive testing.
We aim to recover the bias function between a GIF mock
galaxy catalog and the underlying dark matter catalog.
Since we have both catalogs (obviously not the case with real data), 
we can test our reconstruction against a density-density scatter plot 
based on a cell-by-cell comparison of the two catalogs. Note that
such direct comparison contains significant Poisson scatter 
( apparently the dominant contribution to stochastic bias). Our
procedure recovers the main bias, since it is Poisson noise corrected.

Our SLN3 fits to CPDFs of the GIF galaxy catalog are shown
in Fig.7. The corresponding cumulative PDFs of the galaxies and dark matter
are displayed in Fig. 8. As we have seen in the previous section,
SLN3 is a good approximation for the dark matter distribution.
For the galaxy distribution, SLN3 fits the tail of the distribution
less accurately, even on large scales. As long as
the Poisson kernel is not too broad, we can still use the SLN3 fit
as estimation of PDF for small $\delta$. We can estimate the
limit of the applicability of the SLN3 estimation as
$\delta_{max} \lsim N_{max}/\langle N \rangle -1$, 
with $N_{max}$ being the $N$ where our fit breaks off from measured
tail of the CPDF.

Cumulative PDFs based on the  SLN3 model fit of the mock catalog 
and dark matter are
shown in Fig.8. The bias function directly follows from Equation (1).
We plot $\delta_g$ as function of $\delta_m$ in Fig.9.
This is contrasted with the galaxy density-mass density
scatter plot. Note that errorbars representing the scatter 
due to ``stochastic bias''. The shaded area on the plot
represents simple considerations for the stochasticity of the
bias based on the assumption that all scatter is due to Poisson
noise. It is an excellent approximation to the measured scatter 
(the errorbars),  which appears to show that the scatter is indeed
dominated by Poisson noise. 
We have checked that the measured errorbars are only weakly
dependent on the bin
width $\Delta\delta$, represented by horizontal errorbars:
doubling it produced hardly noticeable effects.
We have repeated the calculations using
Richardson-Lucy method for the galaxies only,
i.e. still SLN3 model for the dark matter. The bias function
is displayed on Figure 9. with dash lines, except for the smallest
scale, where $\langle N \rangle < 0.1$ which we established as the
limit of applicability of this method.

\begin{deluxetable}{crrrrr}
\tablecaption{parameters of SLN3 for the mock galaxy catalog}
\tablewidth{0pt}
\tablehead{
\colhead{Cell Size R ($h^{-1}$Mpc)} &
\colhead{ $\langle N \rangle $ } &
\colhead{ $\langle \log \rho \rangle$ } &
\colhead{ $\sigma_\Phi$ } &
\colhead{ $T_3 \sigma_\Phi$ }&
\colhead{ $T_4 \sigma_\Phi^2$ }
}
\startdata
2.21  & 0.059 & -1.230 & 1.183 & 1.022 & 4.409\\
4.42  & 0.471 & -1.318 & 1.591 & 0.752 & -1.749\\
8.83  & 3.771 & -0.769 & 1.389 & -0.881& 0.902\\
17.66 & 30.166 &-0.250  & 0.798 & -0.869& 2.042\\
 \enddata
\end{deluxetable}

\begin{figure}
\plotone{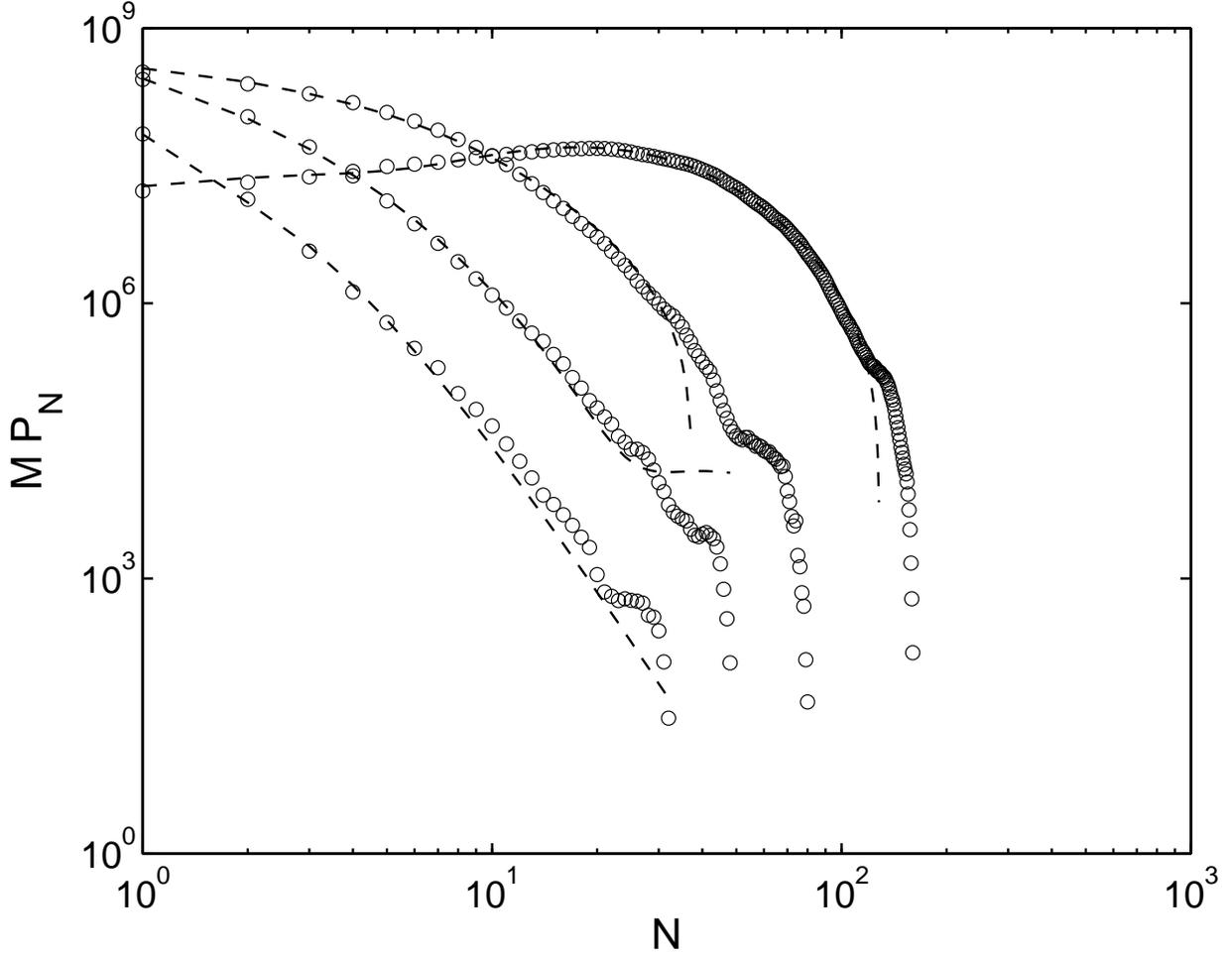}
\caption{The CPDFs of the mock galaxy catalog on scales of
2.21, 4.42, 8.83 and 17.66$h^{-1}$Mpc (from left to right).
Circles are measurements and dash lines are our SLN3 model fits.}
\end{figure}

\begin{figure*}
\resizebox{\hsize}{!}{
\includegraphics{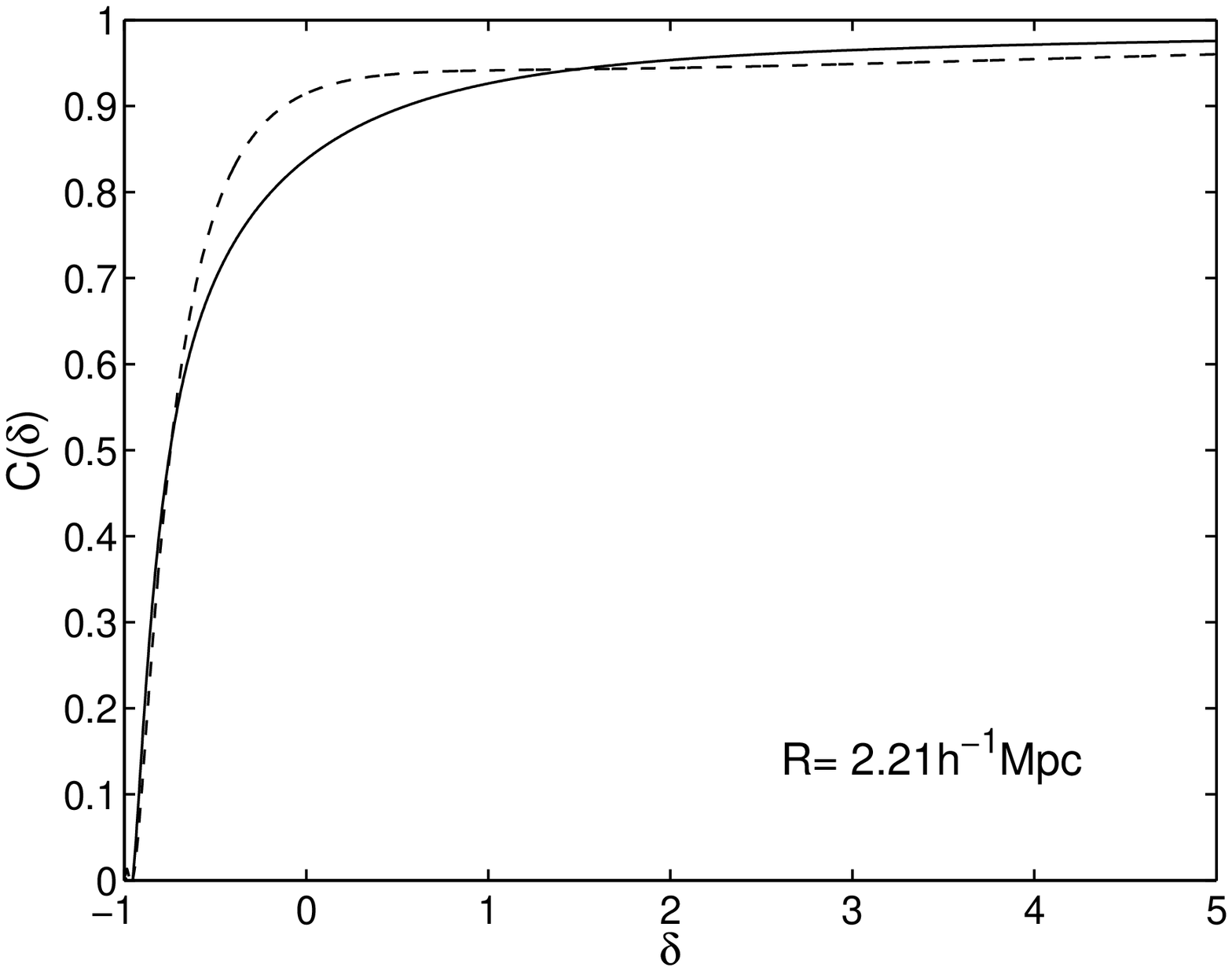}
\includegraphics{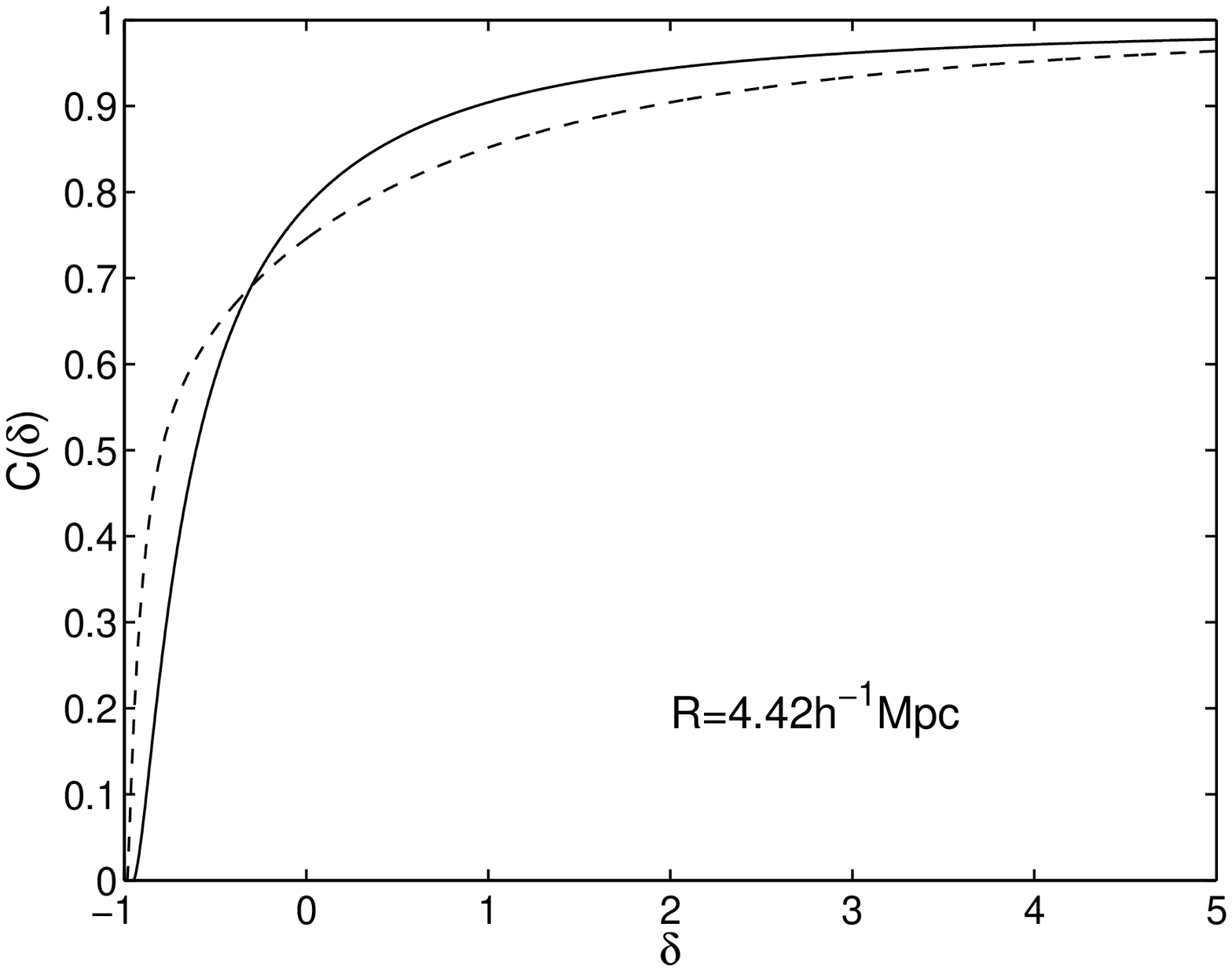} }
\resizebox{\hsize}{!}{
\includegraphics{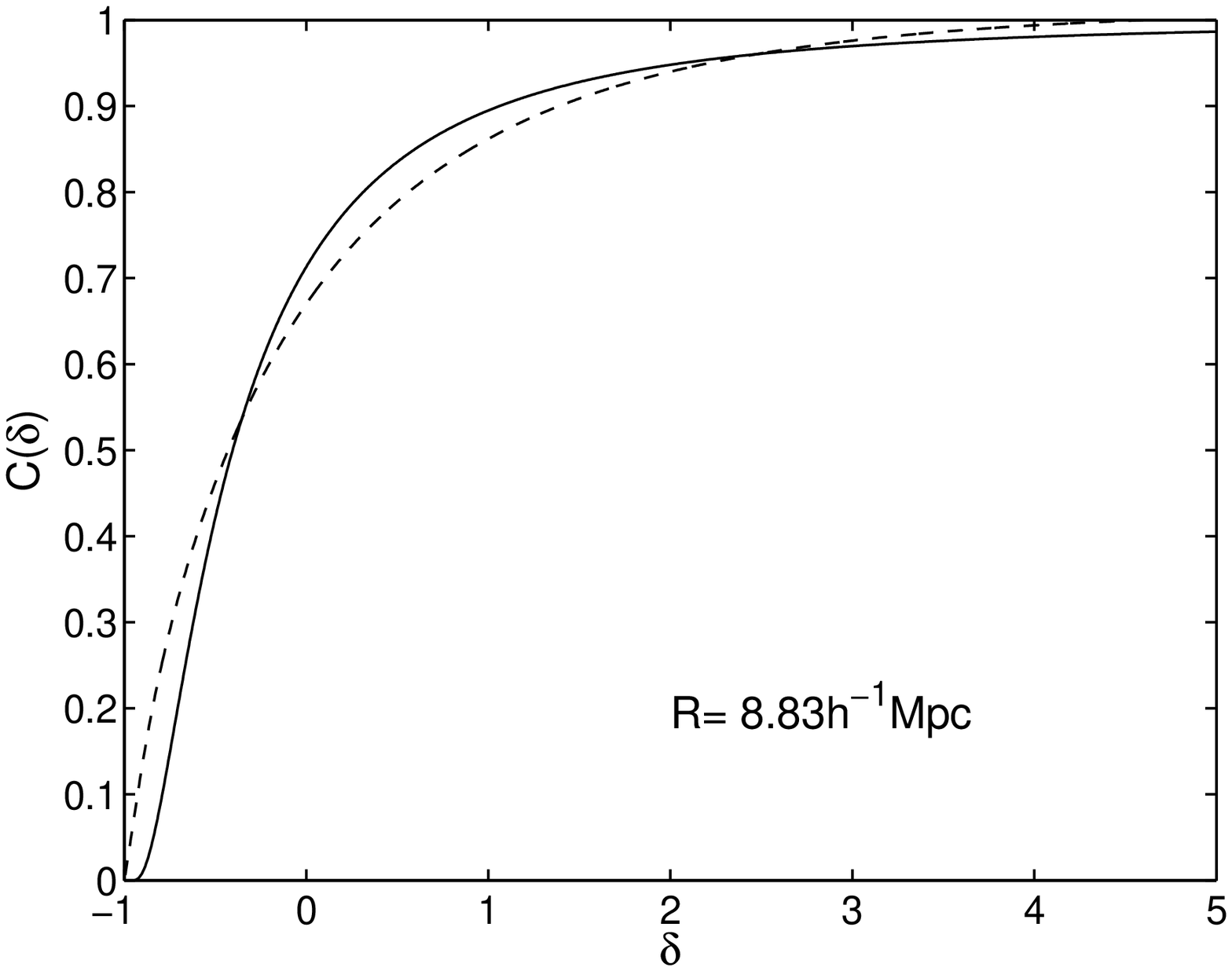}
\includegraphics{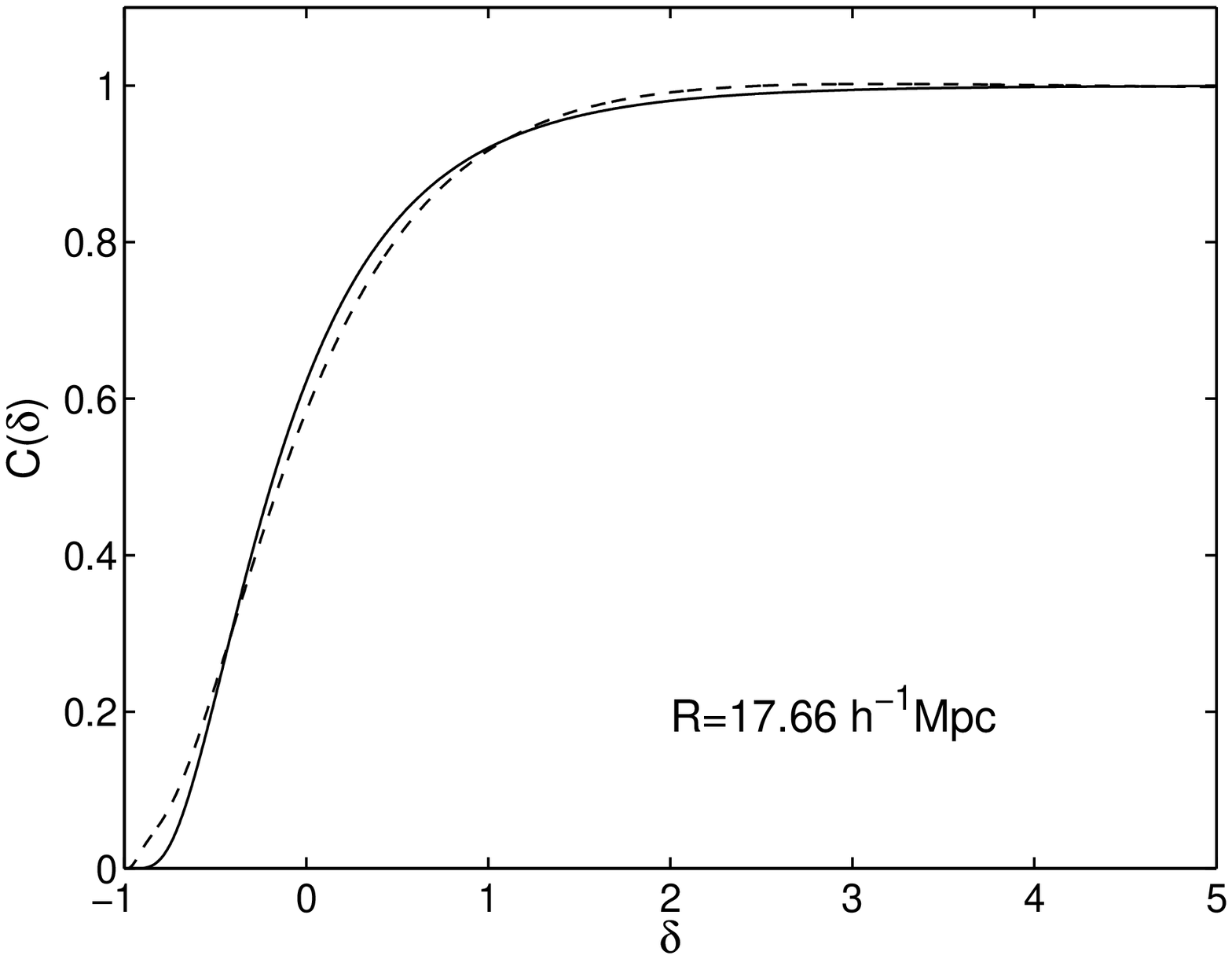} }
\caption{Cumulative PDFs fits based on SLN3 models of 
the mock galaxy catalog
(dashed lines) and dark matter sample (solid lines).}
\end{figure*}

\begin{figure*}
\resizebox{\hsize}{!}{\includegraphics{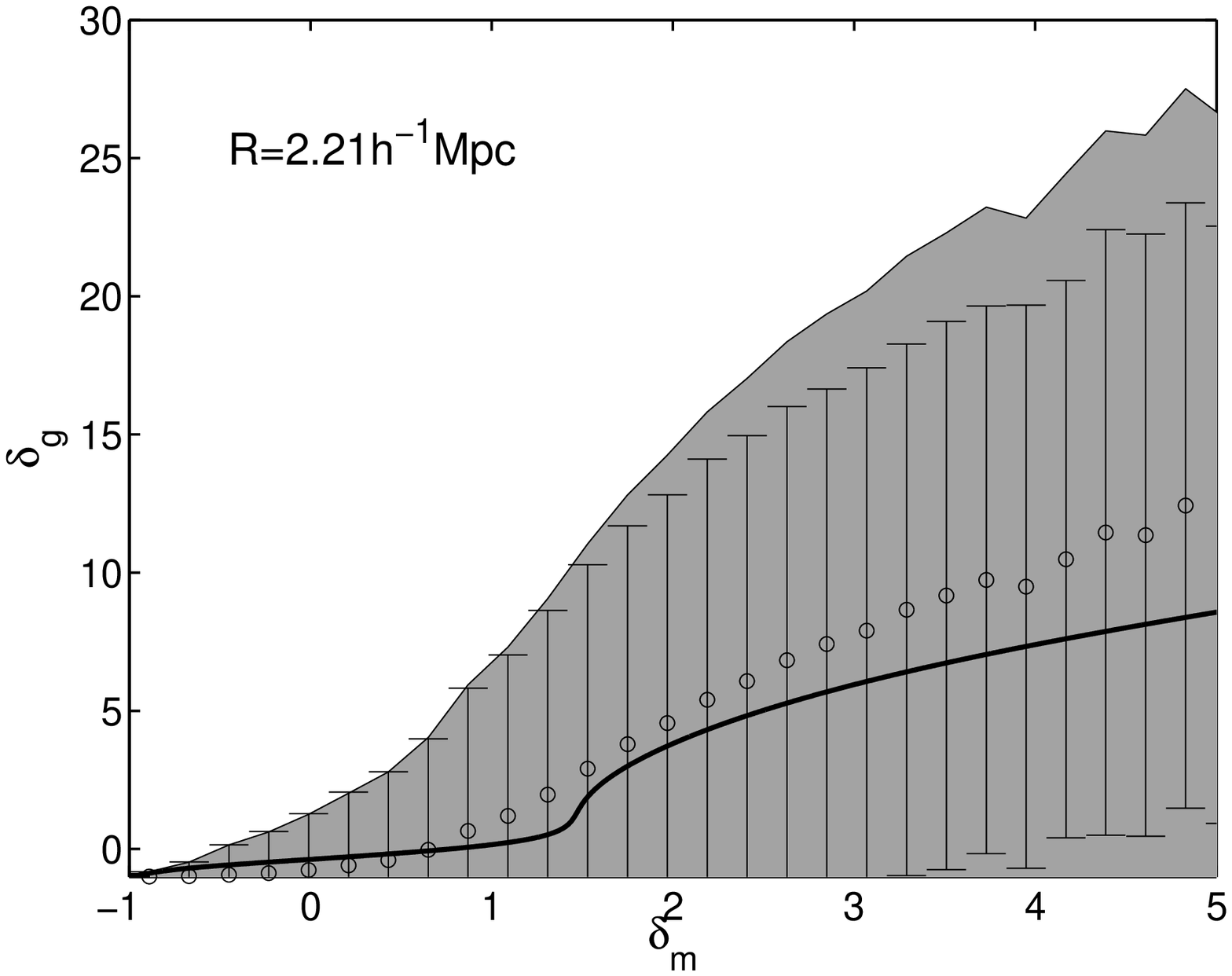}
\includegraphics{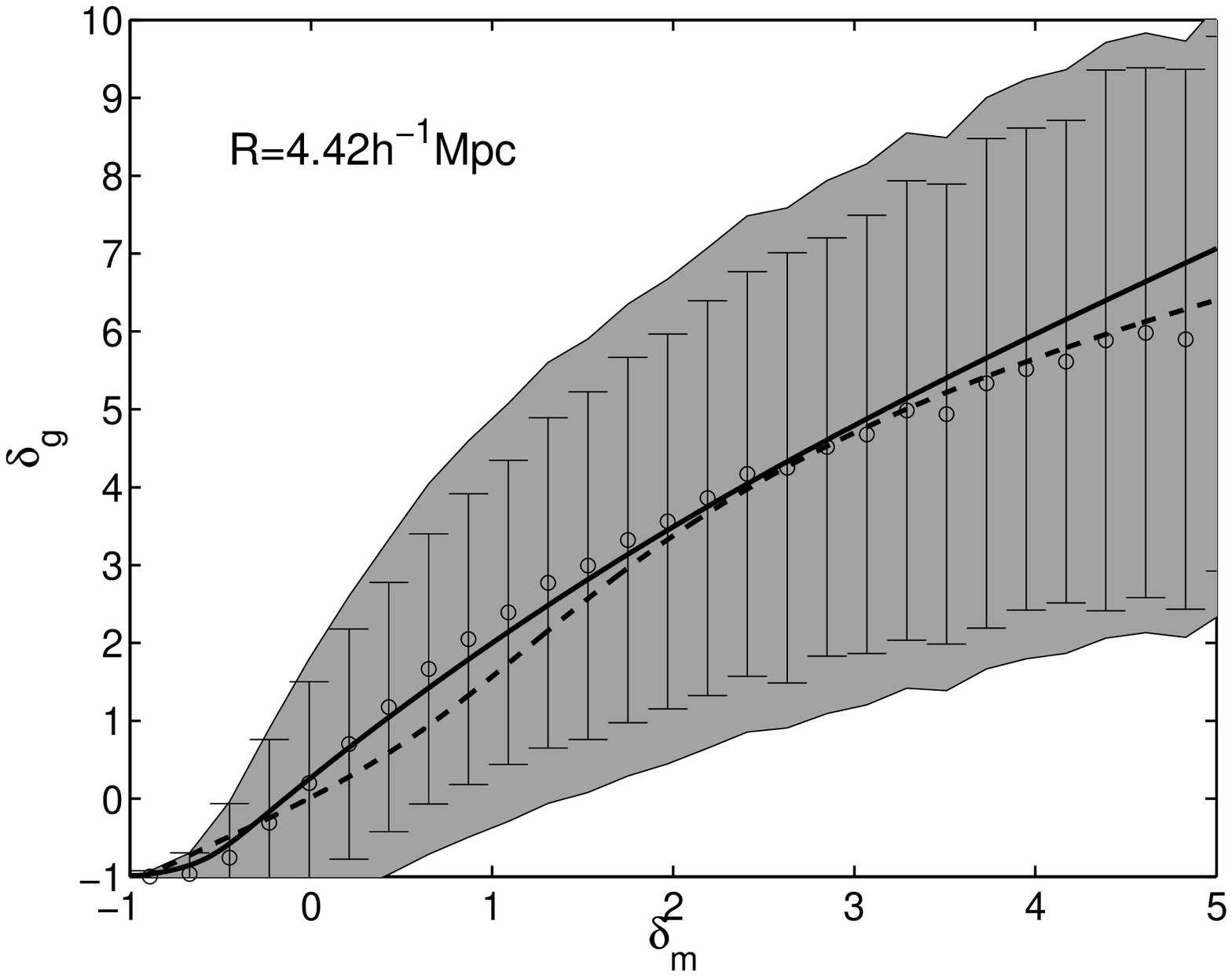}}
\resizebox{\hsize}{!}{\includegraphics{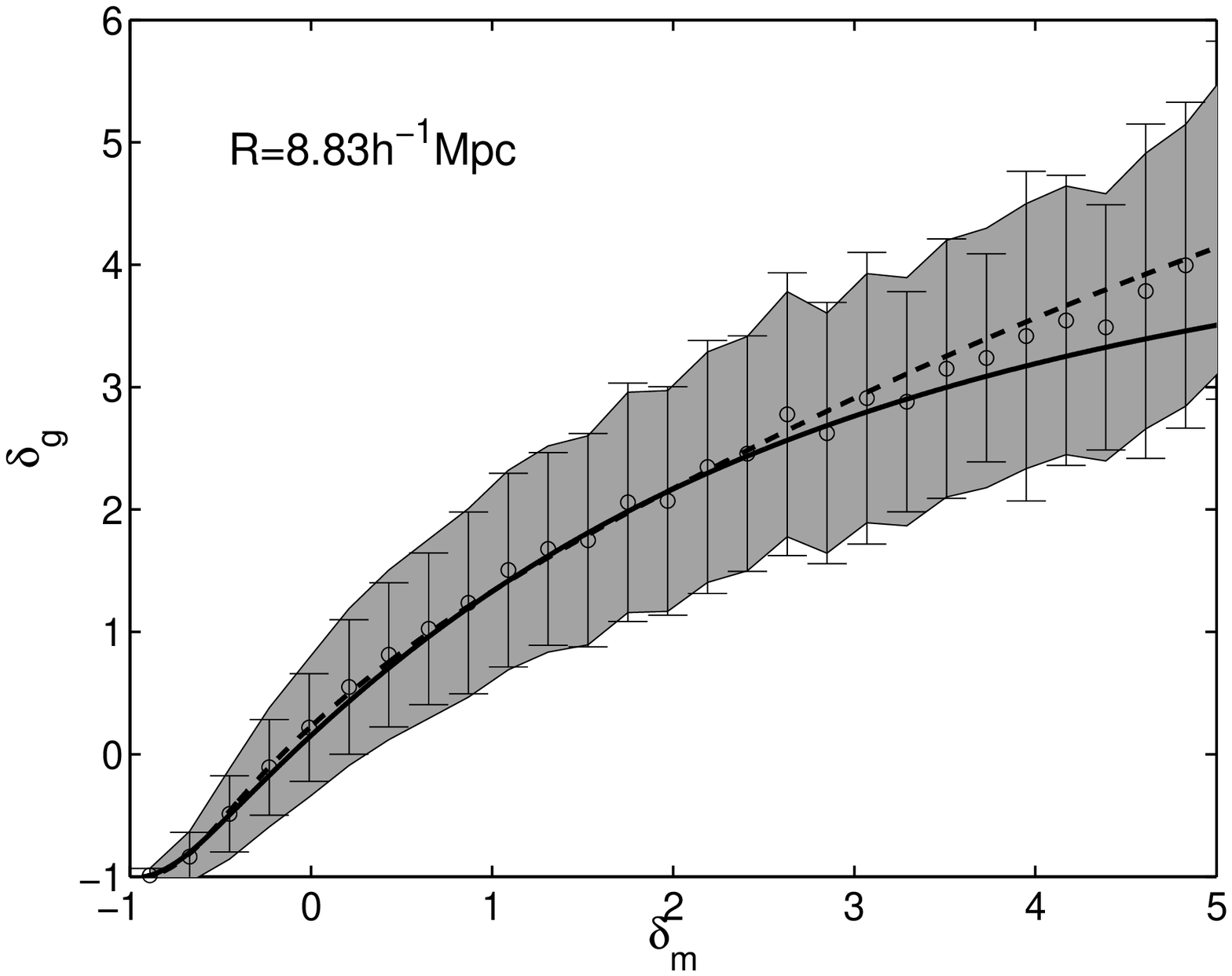}
\includegraphics{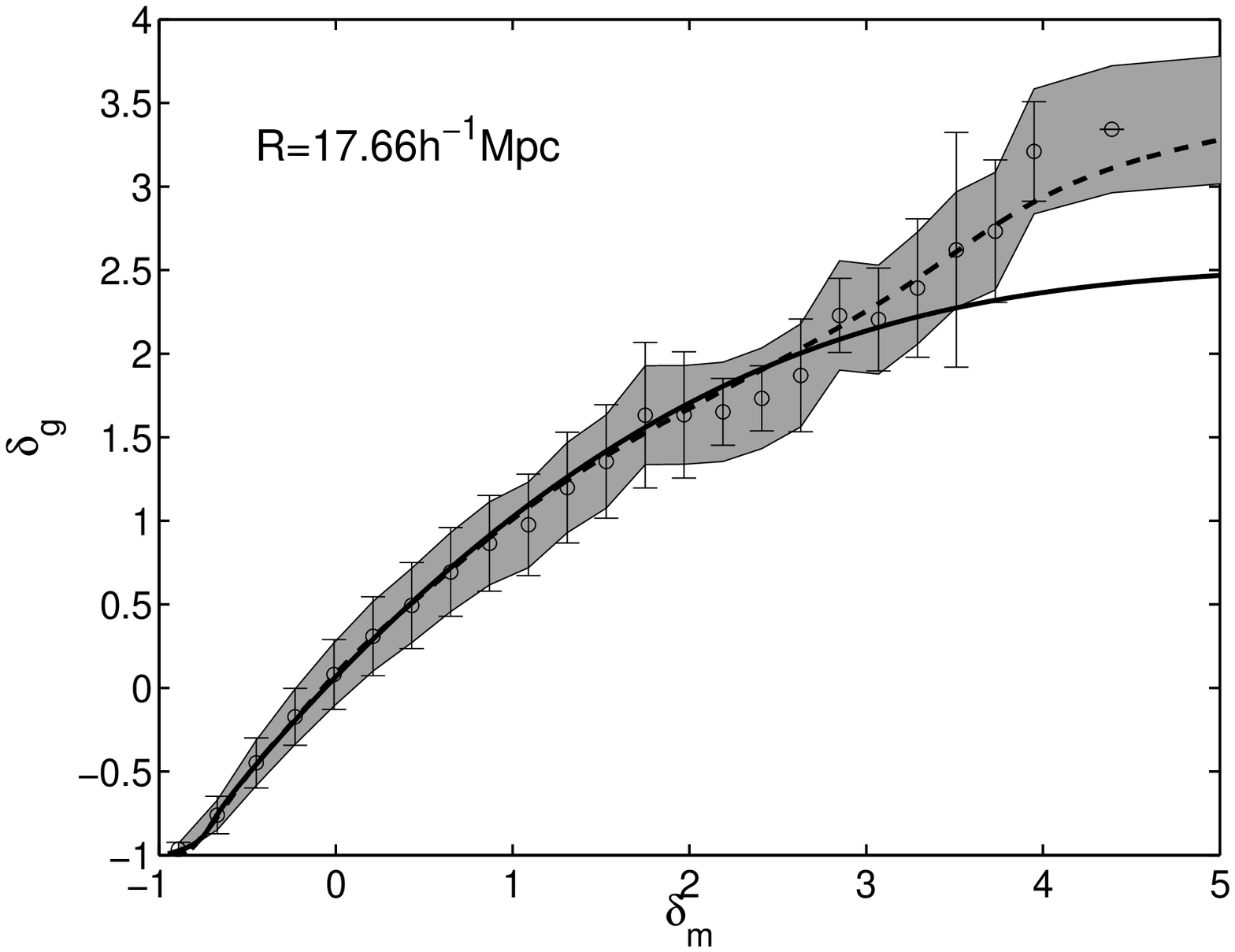}}
\caption{
Bias function $\delta_g=f(\delta_m)$ for the mock galaxy catalog.
Solid lines are from the SLN3 model galaxy PDFs, 
dash lines are based on the
Richardson-Lucy method. The reference mass PDFs were obtained
from an SLN3 fit.
Circles are the $\langle \delta_g | \delta_m \rangle$ measured directly from
samples, errorbars show  their $1\sigma$ scatter. The full 
width of $\delta_m$ bins of the scatter plot is $\Delta\delta_m=0.22$.
The shaded area represents a simple Poisson scatter, which appears
to be an excellent approximation to the stochasticity of the bias
in these simulations.}
\end{figure*}

\subsection{Bias from the GIF Mock Galaxy Catalog for SDSS}

This mock catalog contains more galaxies than the previous GIF mock
galaxy catalog to match the density of the SDSS.
The fits with SLN3 to its CPDFs are shown on Fig.10. 
Fitted parameters of the model are listed in Table 3. Bias functions are 
displayed in Fig.11.

\begin{figure}
\plotone{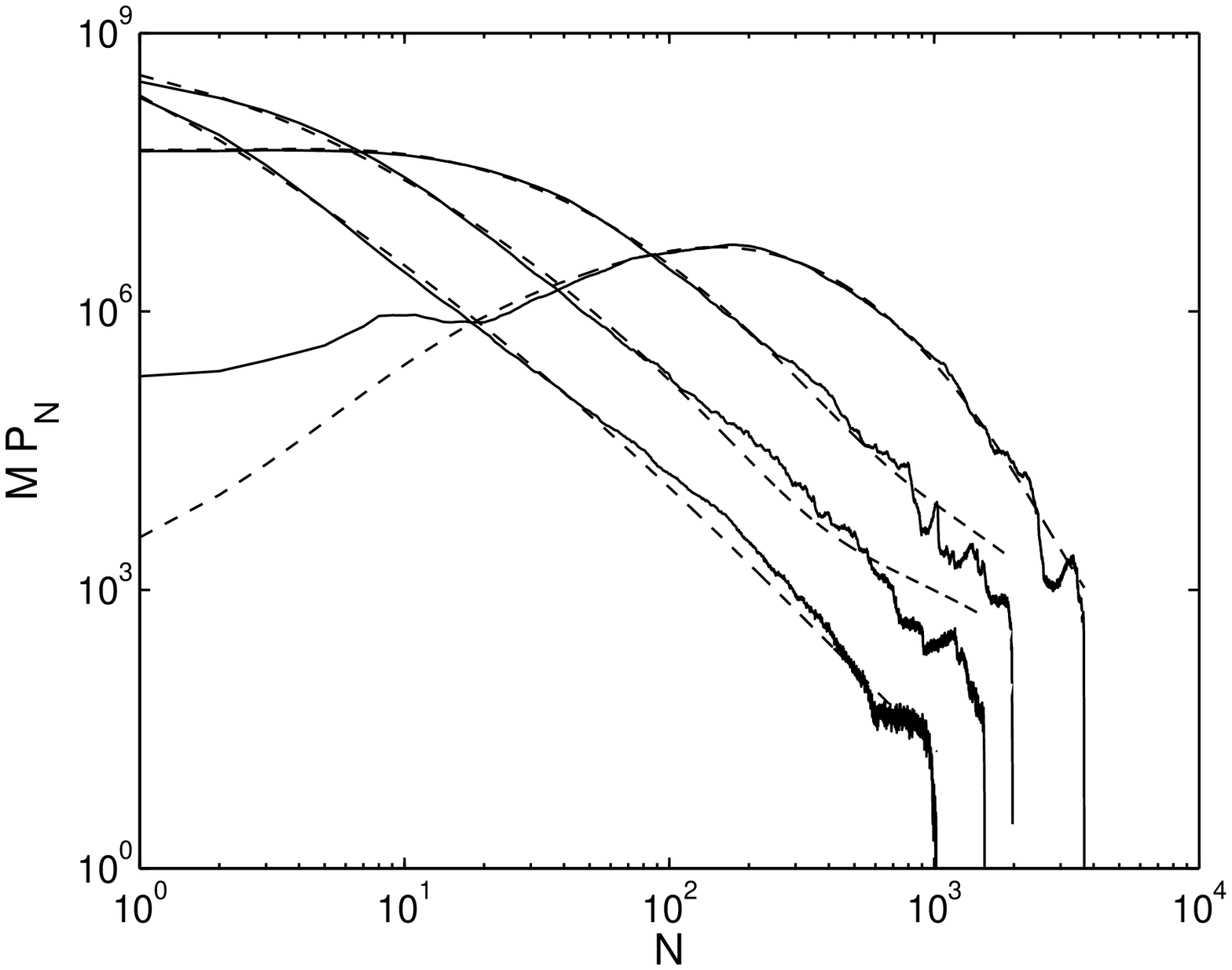}
\caption{The CPDFs of the GIF mock galaxy catalog for SDSS on scales of
2.21, 4.42, 8.83 and 17.66$h^{-1}$Mpc (from left to right). Solid lines are
from CIC measurements, dash lines are from SLN3 fit.}
\end{figure}

\begin{deluxetable}{crrrrr}
\tablecaption{parameters of SLN3 for the mock galaxy catalog for SDSS}
\tablewidth{0pt}
\tablehead{
\colhead{Cell Size R ($h^{-1}$Mpc)} &
\colhead{ $\langle N \rangle $ } &
\colhead{ $\langle \log \rho \rangle$ } &
\colhead{ $\sigma_\Phi$ } &
\colhead{ $T_3 \sigma_\Phi$ }&
\colhead{ $T_4 \sigma_\Phi^2$ }
}
\startdata
2.21  & 0.7198 & -2.446 & 2.167 & 0.181 & -0.300 \\
4.42  & 5.758  & -1.631 & 1.971 & -0.387 & 0.557 \\
8.83  & 46.064 & -0.796 & 1.347 & -0.338& 0.968\\
17.66 & 368.514 &-0.318  & 0.824 & -0.231& 0.280\\
\enddata
\end{deluxetable}

\begin{figure*}
\resizebox{\hsize}{!}
{\includegraphics{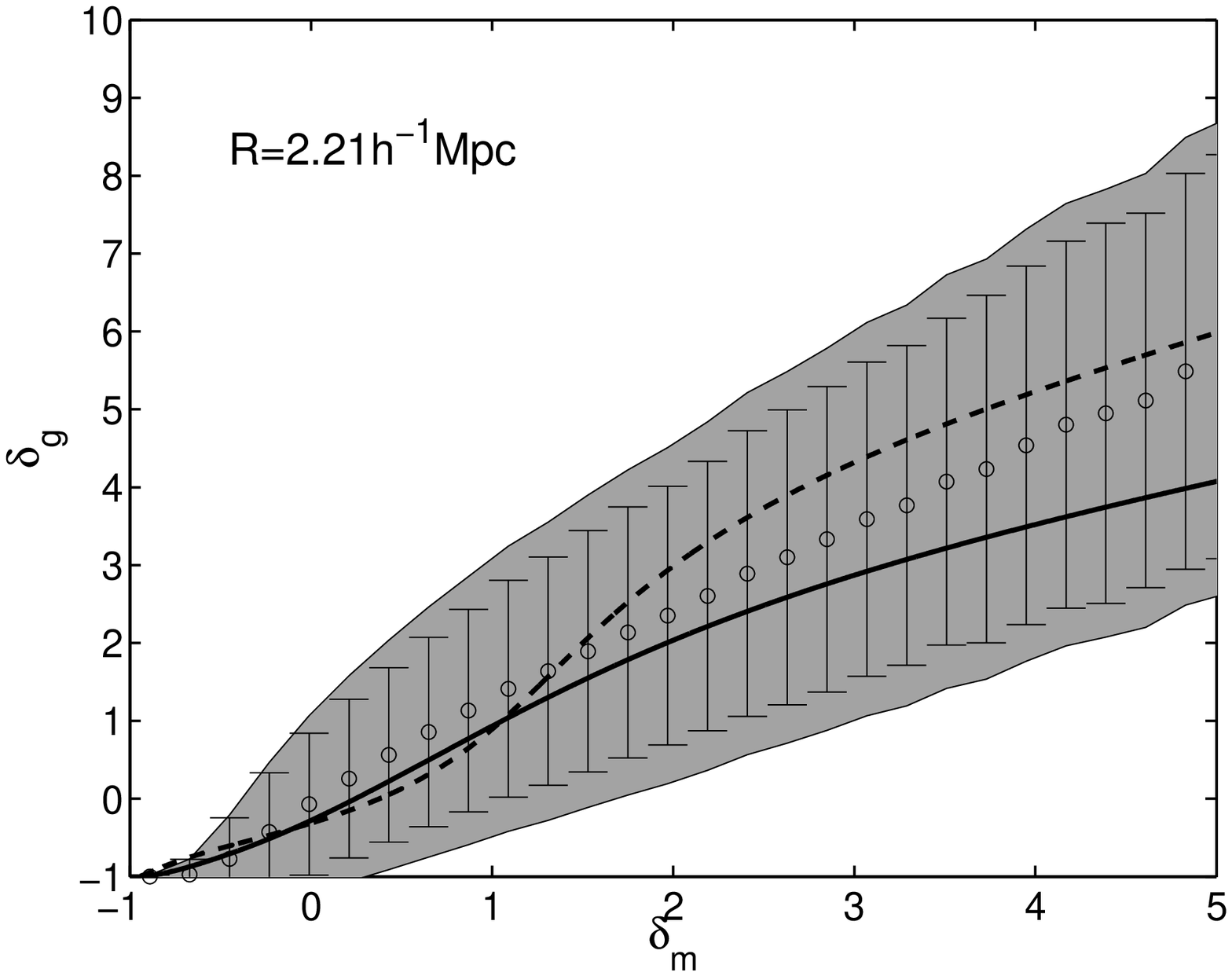}\includegraphics{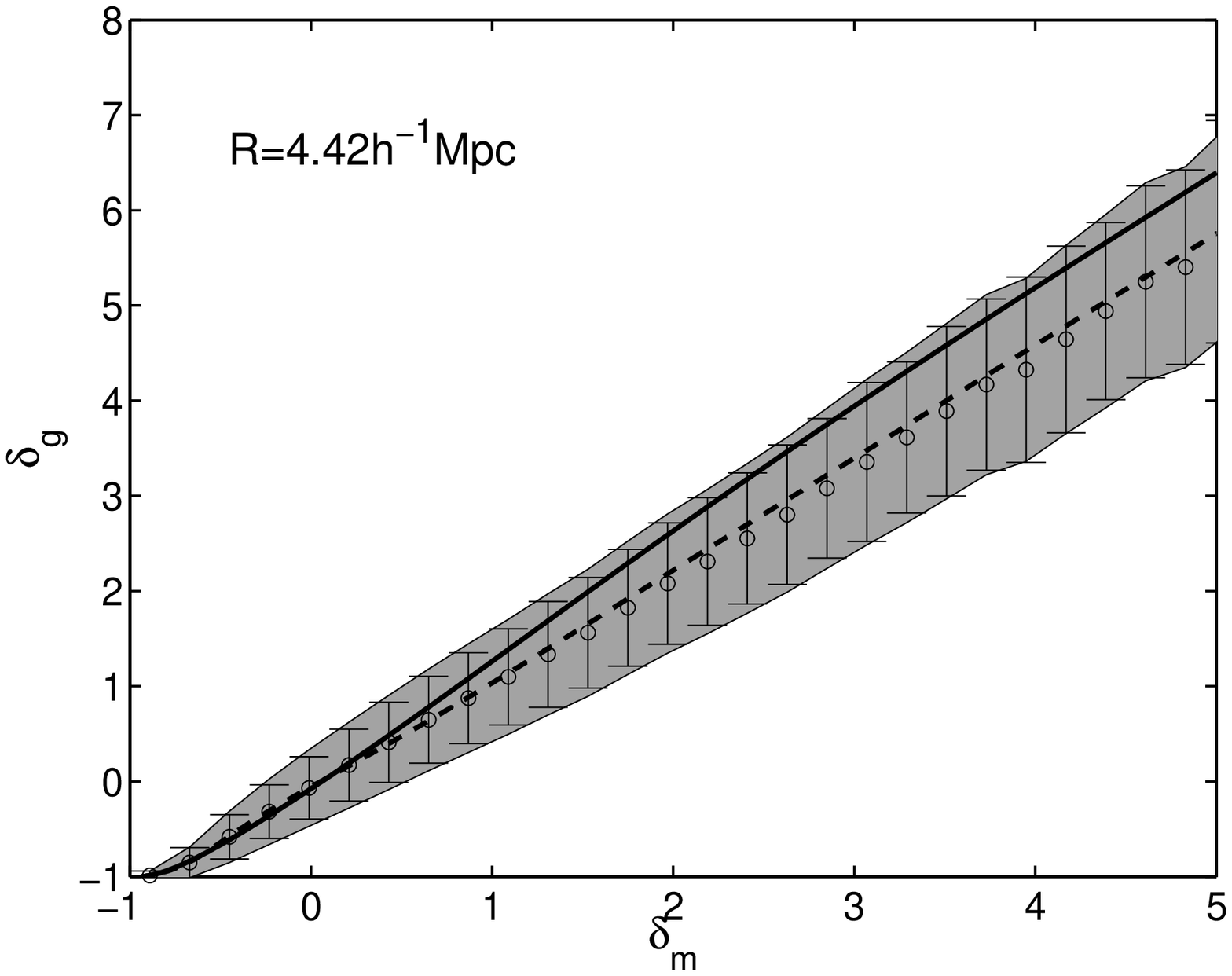}}
\resizebox{\hsize}{!}
{\includegraphics{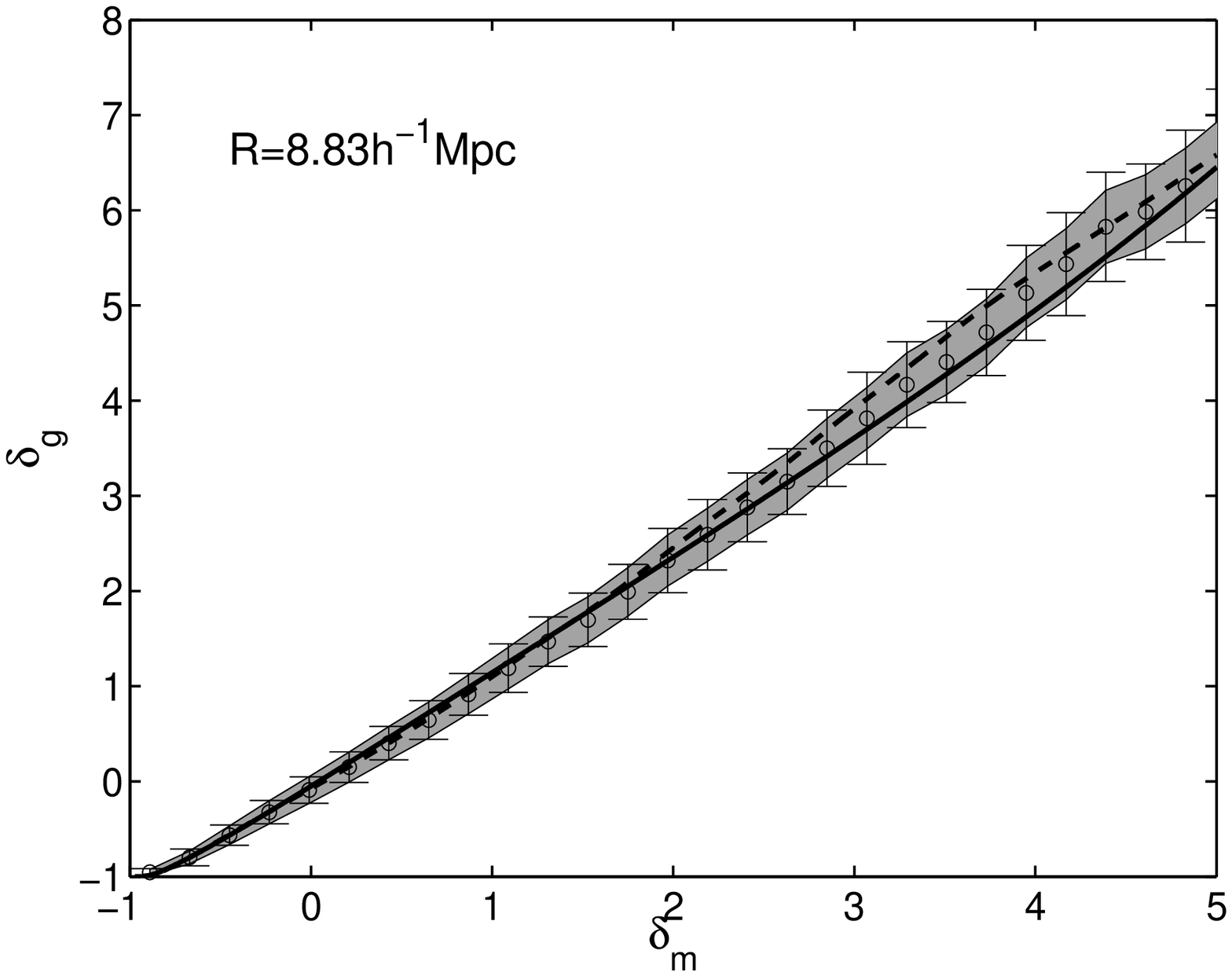}\includegraphics{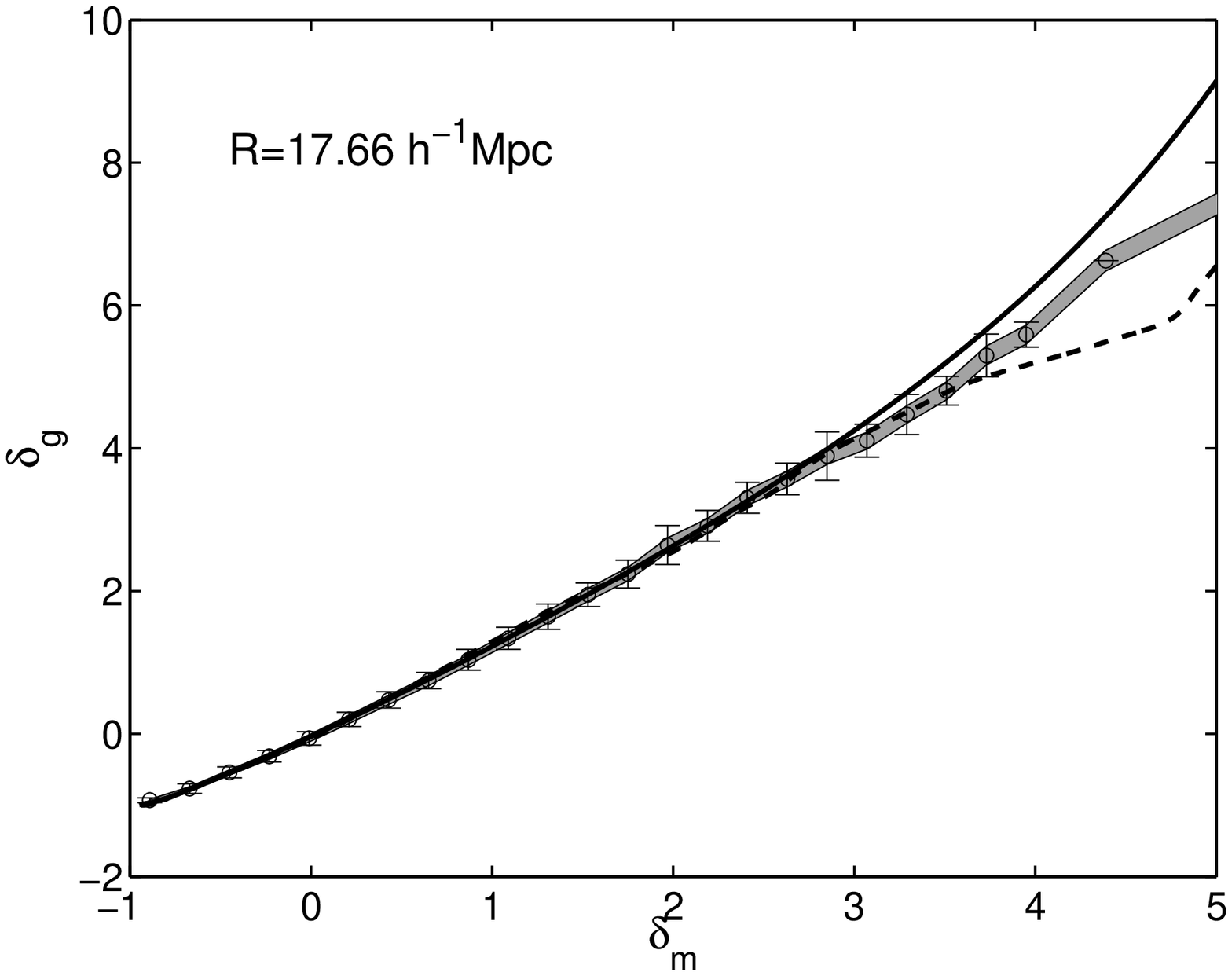}}
\caption{Same as Figure 9 but for the SDSS mock
galaxy catalog.}
\end{figure*}

\subsection{Redshift space}

All the tests we have done so far in real space 
have been repeated in redshift space.
The plausible assumption about the application of our method in 
redshift space is that, in absence of significant velocity bias,
the redshift space CPDFs will be modified similarly in the
dark matter and galaxy catalogs. As long as this is correct, the bias function
can be obtained by direct application of the method in redshift space.
However, even if this assumption is only approximate, the dark matter
distribution in redshift space is still recovered, which means
that the effects of galaxy formation are decoupled from the
evolution of dark matter; this is the principal goal of bias
recovery. The tests in this section show that redshift distortions
have a small effect on our procedure, thus our recipe can be safely
applied to redshift surveys without significant corrections. It has
to be born in mind though, that this statement is somewhat model
dependent; we assume that the simulations and the semianalytic models
used to create mock catalogs are close enough to reality that this
statement will hold for real data.

The CPDFs and their SLN3 fits for the dark matter sample, the GIF mock galaxy 
catalog and the mock galaxy catalog for SDSS are shown
in Fig.12, Fig.13 and  Fig.14 respectively. The
recovered bias functions of the GIF mocks galaxy are displayed 
in Fig.15, while for the SDSS mocks in Fig.16. 

The recovery of bias in redshift space is just as successful as in
real space. We have also compared the recovered red shift space bias
function to that of the real space. Results based on RL inversion only
are shown on Figure 17 (GIF galaxy mock catalog) and Figure 18 (mock galaxy
catalog for SDSS). The difference between the two curves is generally
small, in fact for $\delta_m \lsim 2$, the effect of
redshift distortions on the bias function is negligible. The 
apparent difference between the GIF curves on 
scale of $4.42h^{-1}$Mpc is mostly due to
the fact that the RL inversion is pushed to its limits
at $\langle N \rangle=0.47$.

\begin{figure}
\plotone{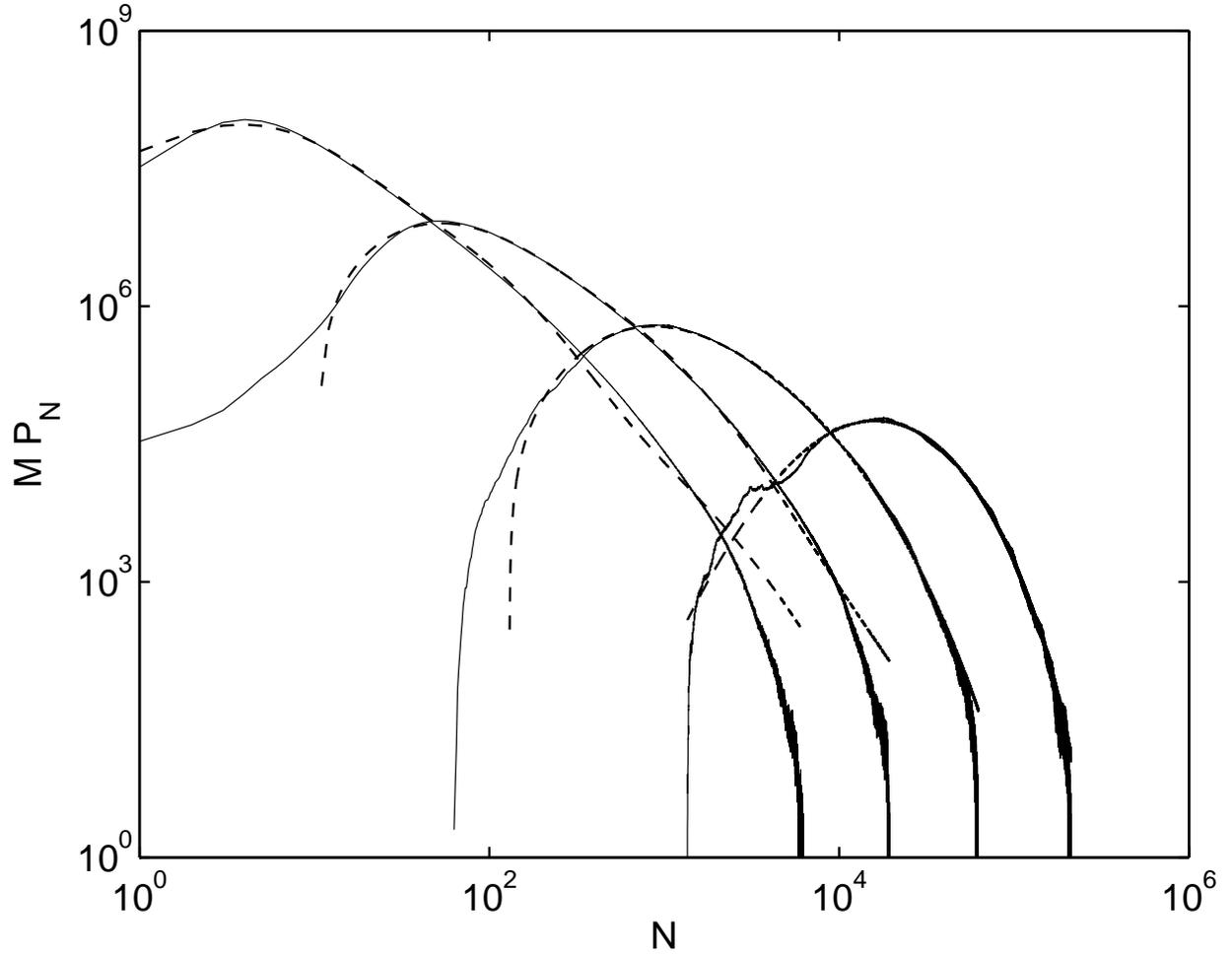}
\caption{CPDFs and their SLN3 model fits for the dark matter in redshift 
space. Solid lines correspond to the  measurements, while 
dash lines show SLN3 model fits.
For these curves from left to right, 
cells size at which CPDFs are measured are
$R=2.21, 4.42, 8.83, 17.66h^{-1}$Mpc.}
\end{figure}

\begin{figure}
\plotone{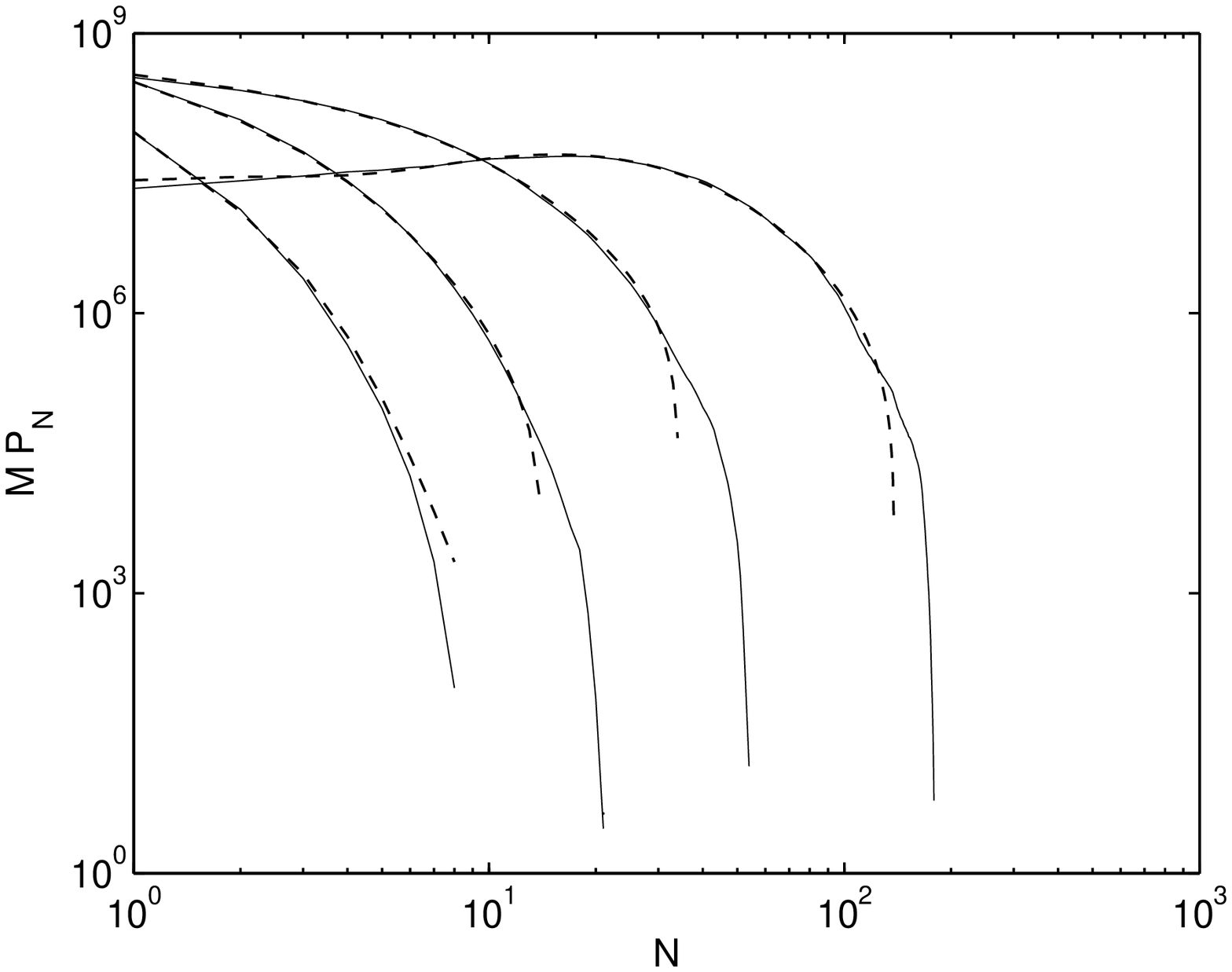}
\caption{Same as Figure 12. but for the the GIF mock galaxy catalog
in redshift space.}
\end{figure}

\begin{figure}
\plotone{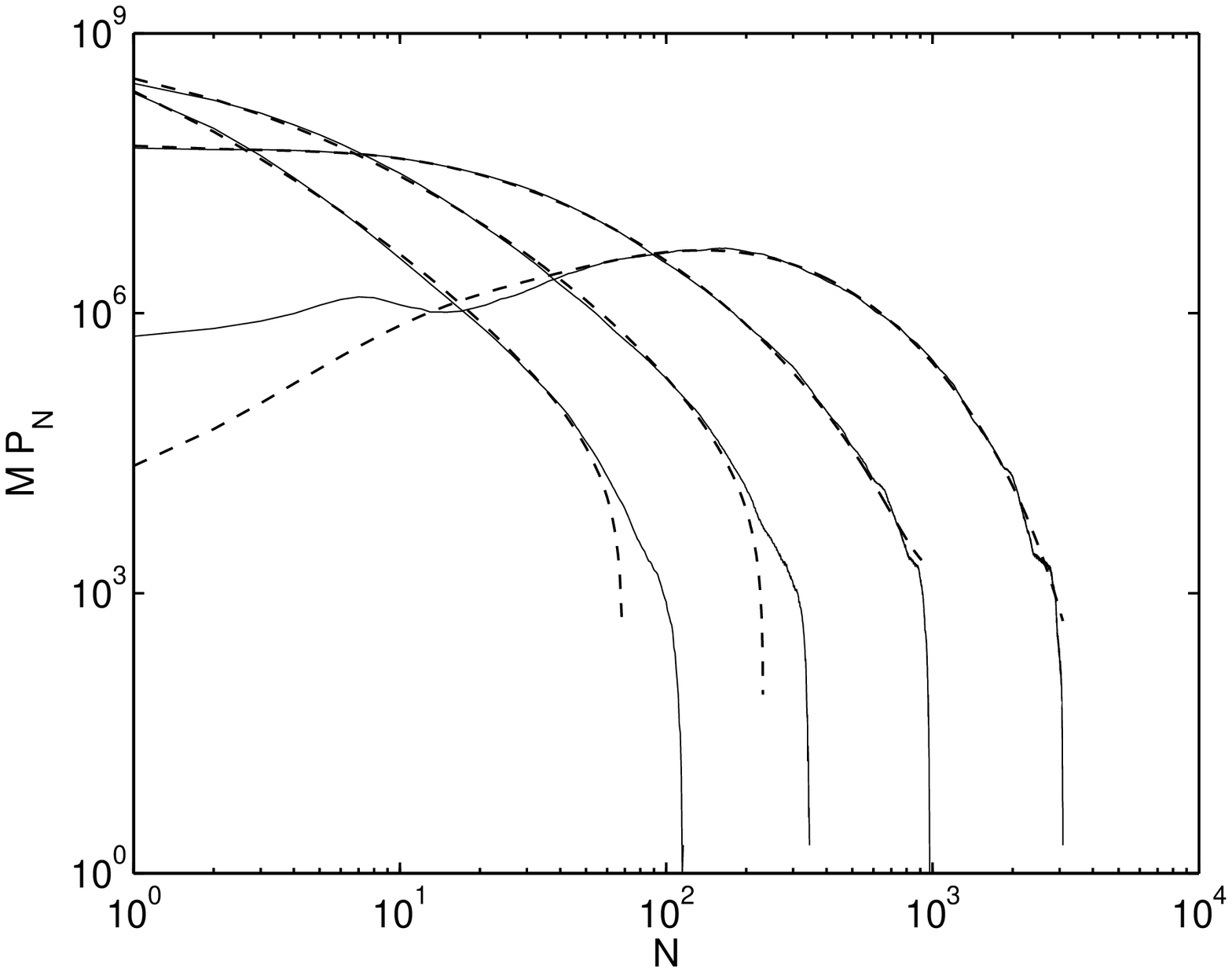}
\caption{Same as Figure 12. but for the GIF mock galaxy catalog
for the SDSS in redshift space}
\end{figure}

\begin{figure*}
\resizebox{\hsize}{!}
{\includegraphics{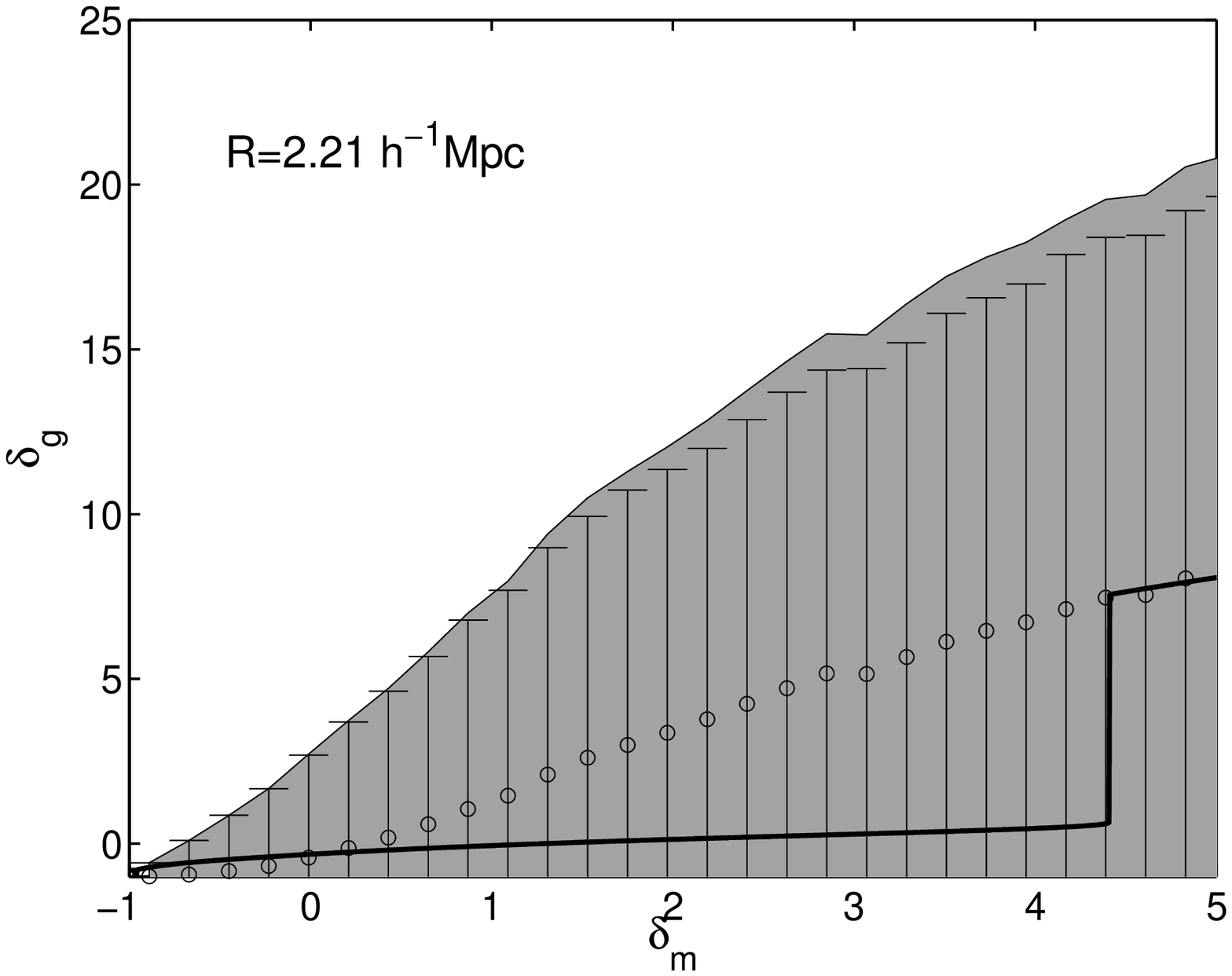}\includegraphics{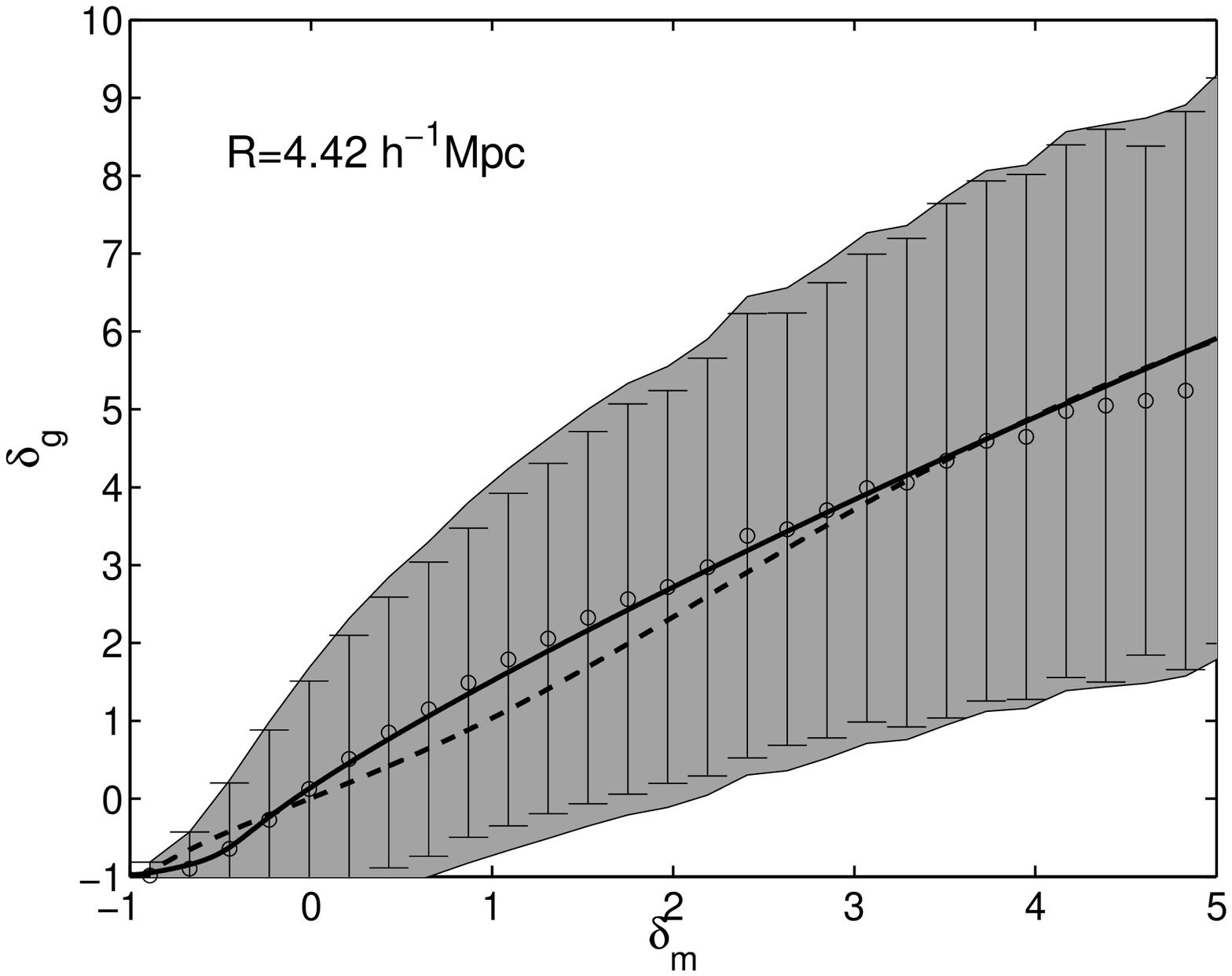}}
\resizebox{\hsize}{!}
{\includegraphics{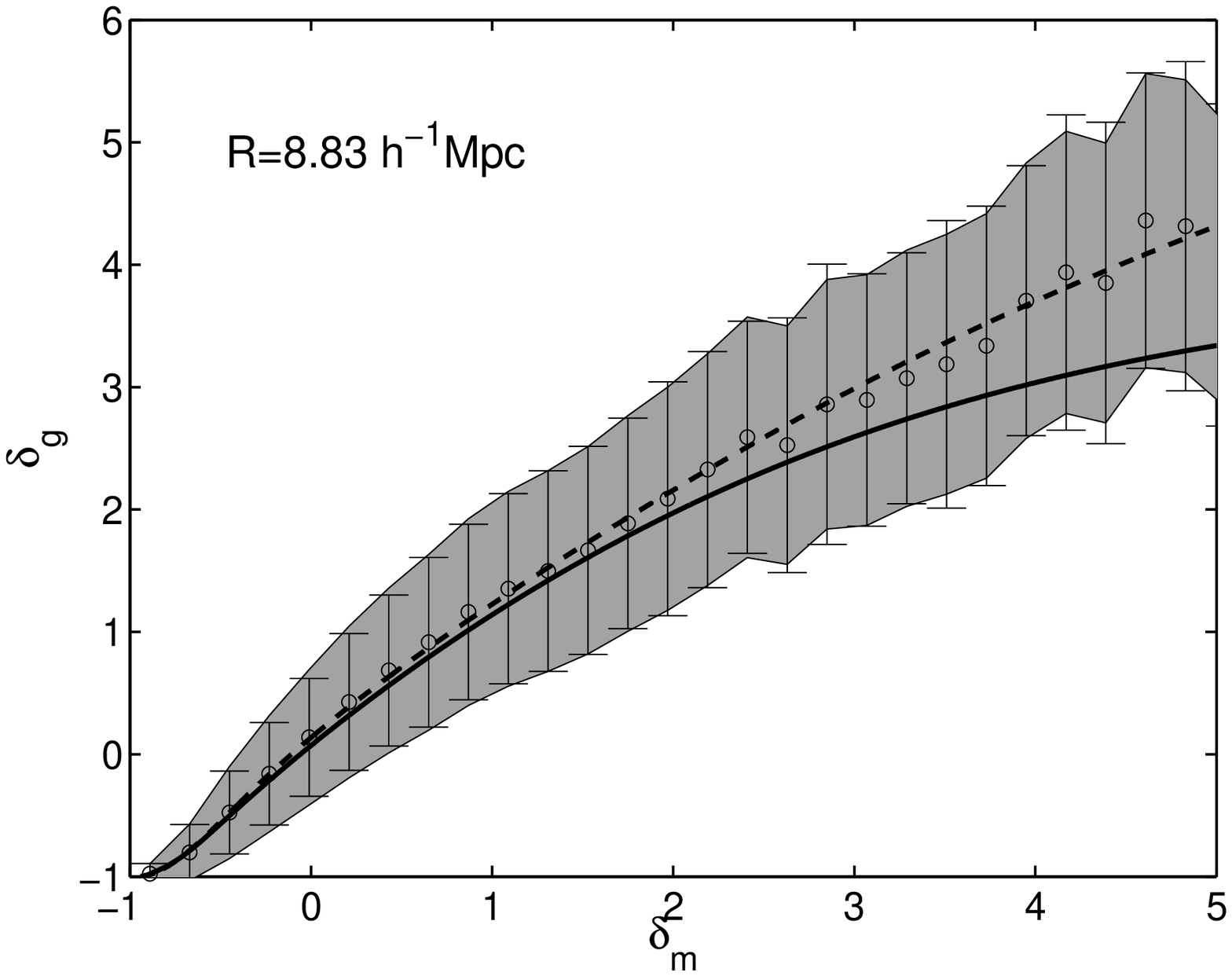}\includegraphics{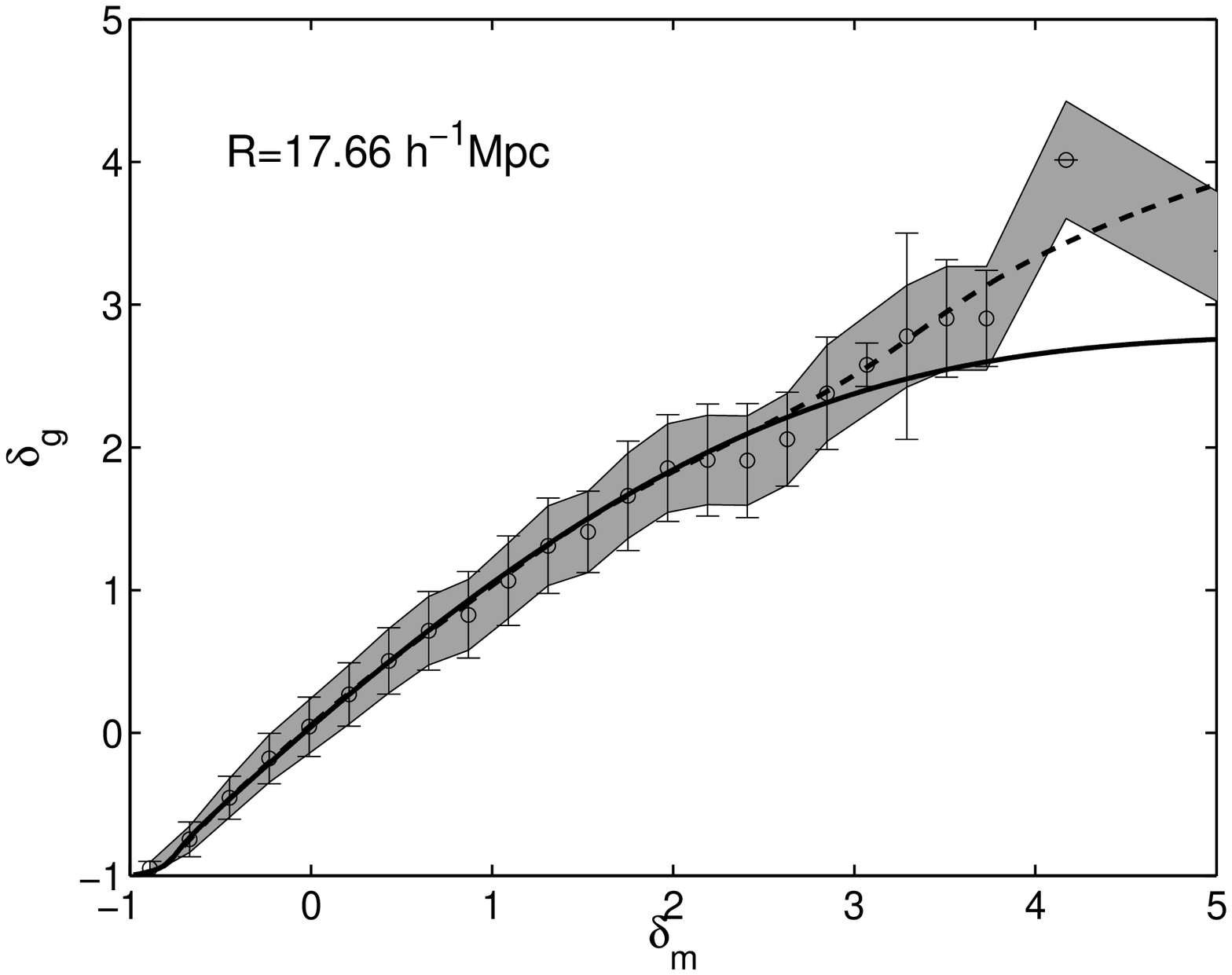}}
\caption{Same as Figure 9. but for the GIF
galaxy catalog in redshift space.}
\end{figure*}

\begin{figure*}
\resizebox{\hsize}{!}
{\includegraphics{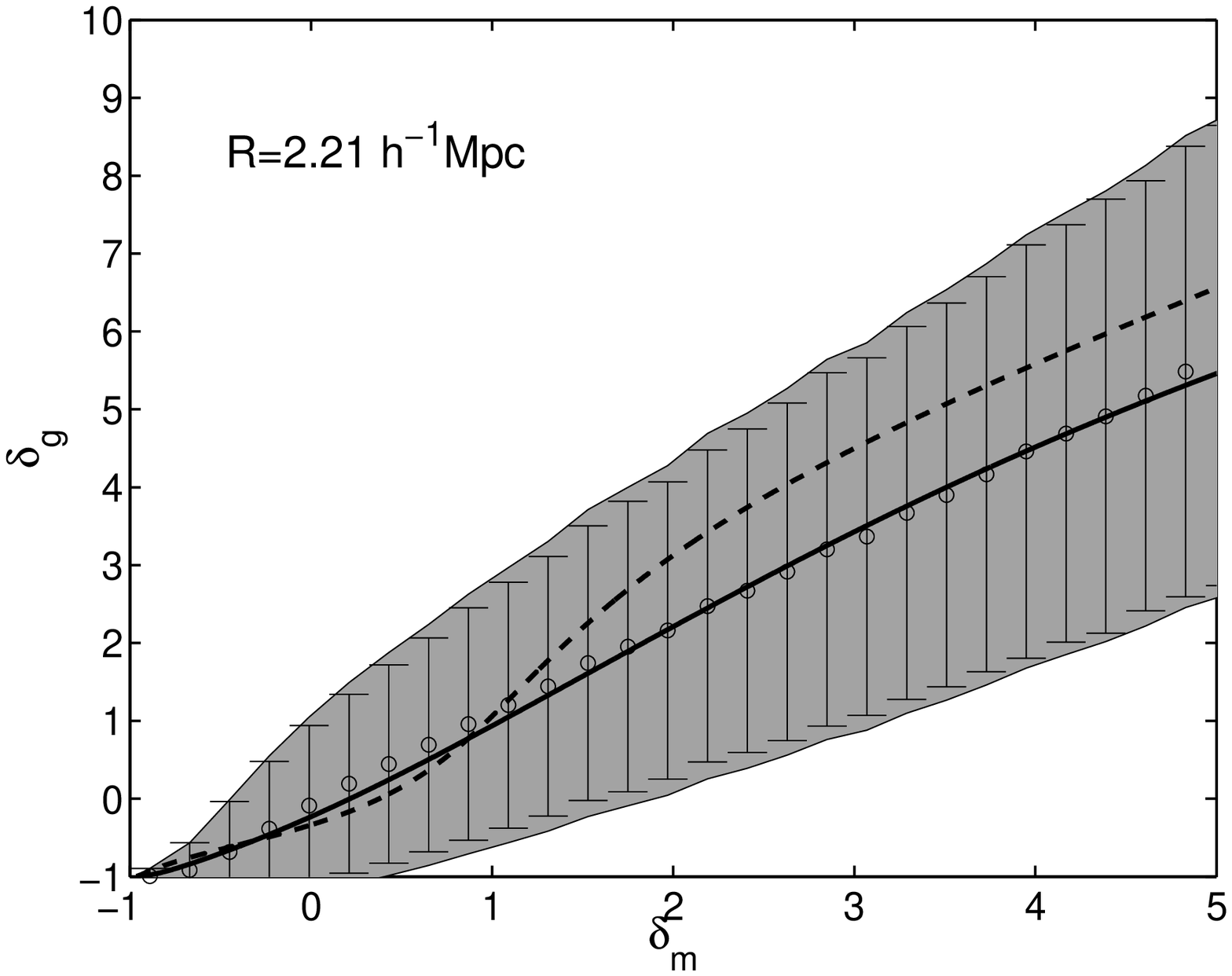}\includegraphics{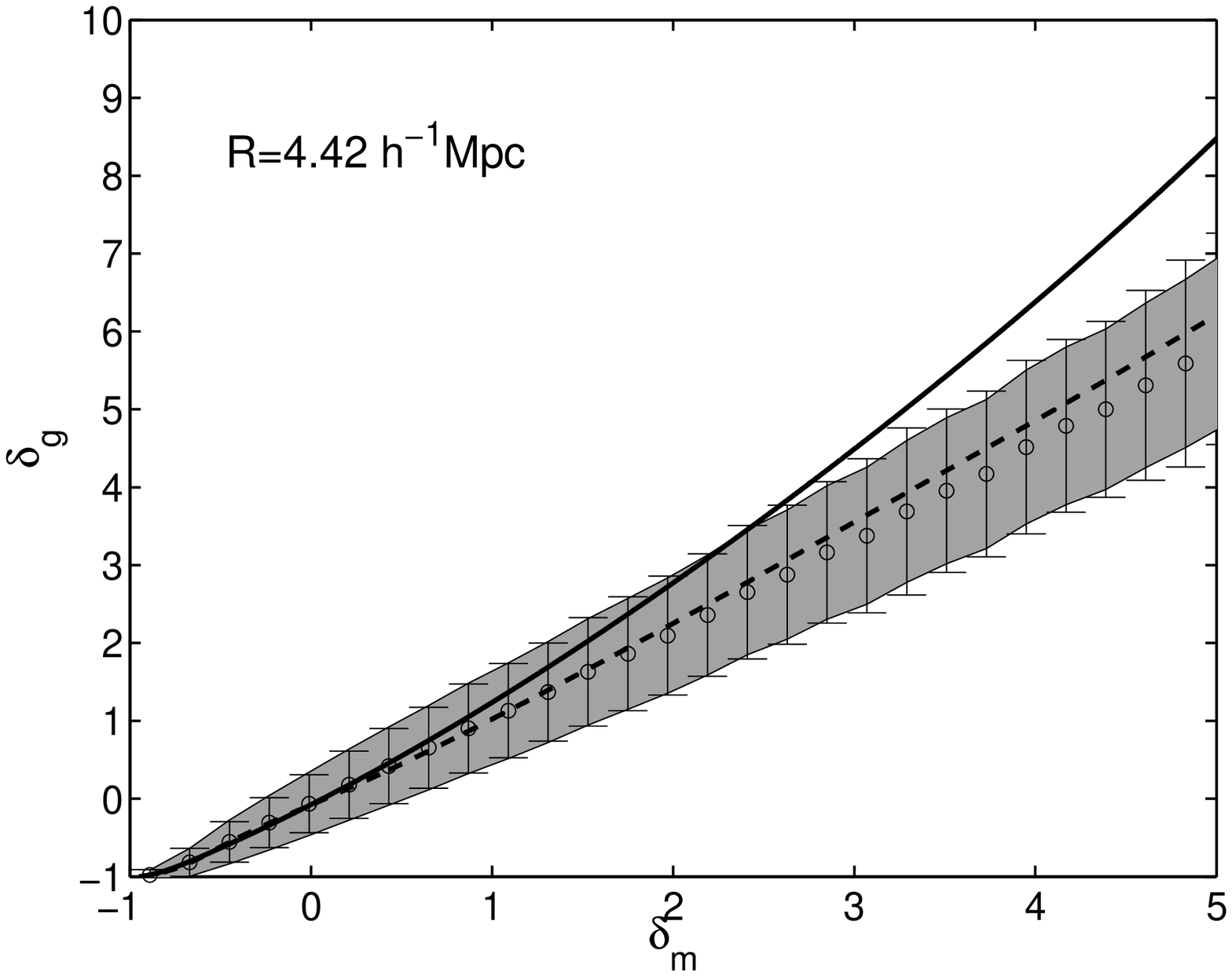}}
\resizebox{\hsize}{!}
{\includegraphics{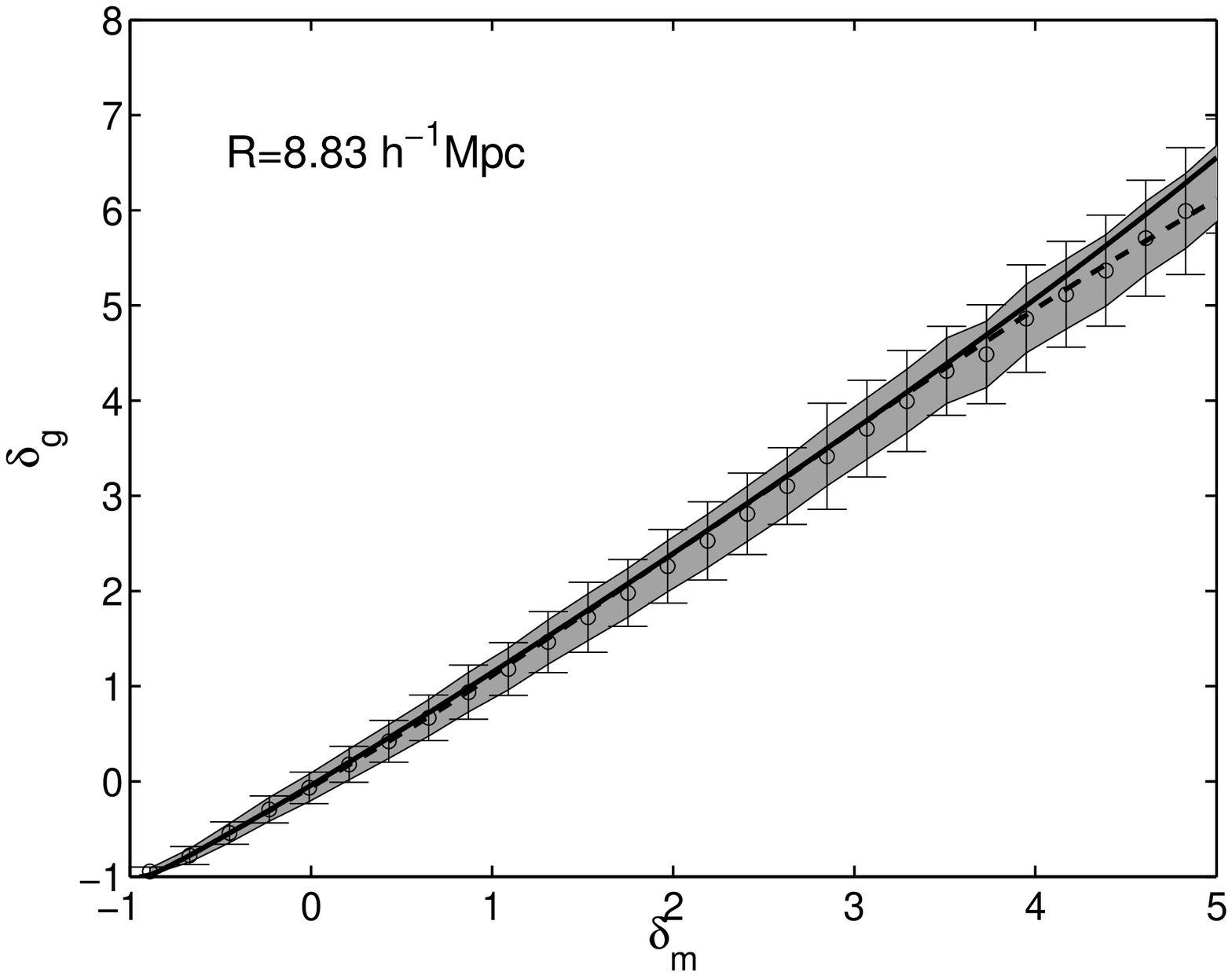}\includegraphics{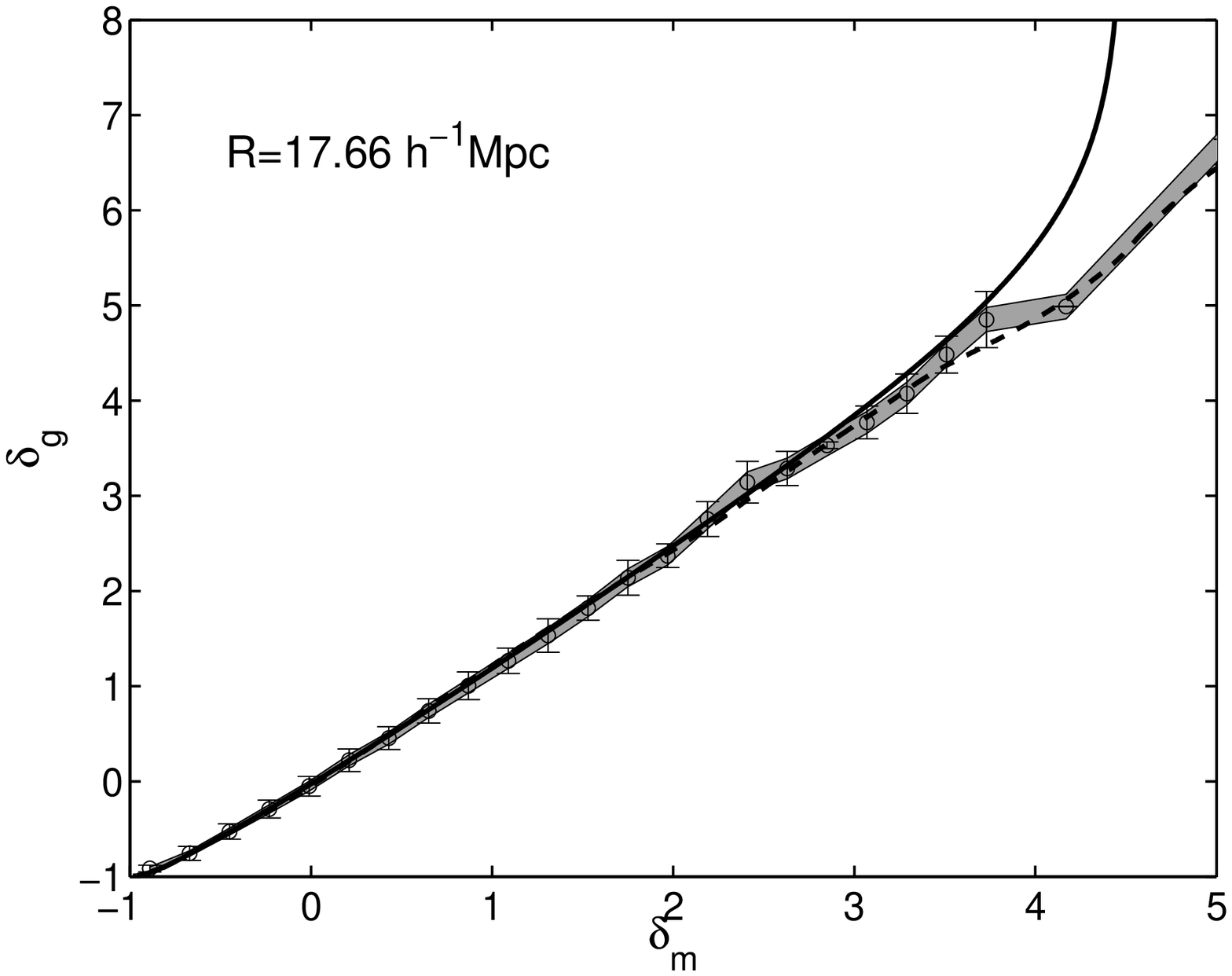}}
\caption{Same as Figure 9. but for the GIF mock galaxy catalog 
for the SDSS in redshift space}
\end{figure*}

\begin{figure}
\plotone{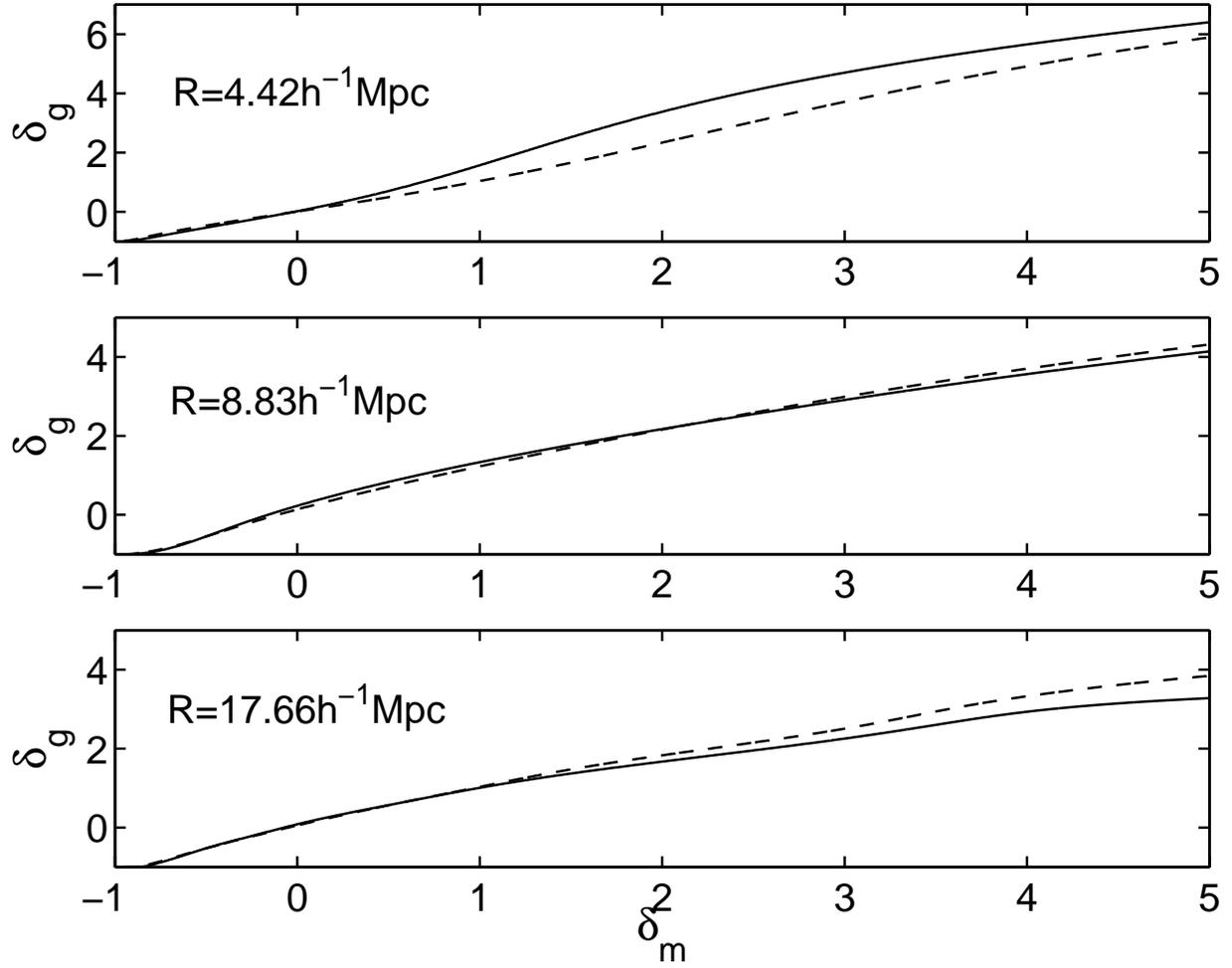}
\caption{Comparison of recovered bias functions of the GIF galaxy catalog
in real space (solid lines) and in redshift space (dashed lines).}
\end{figure}

\begin{figure}
\plotone{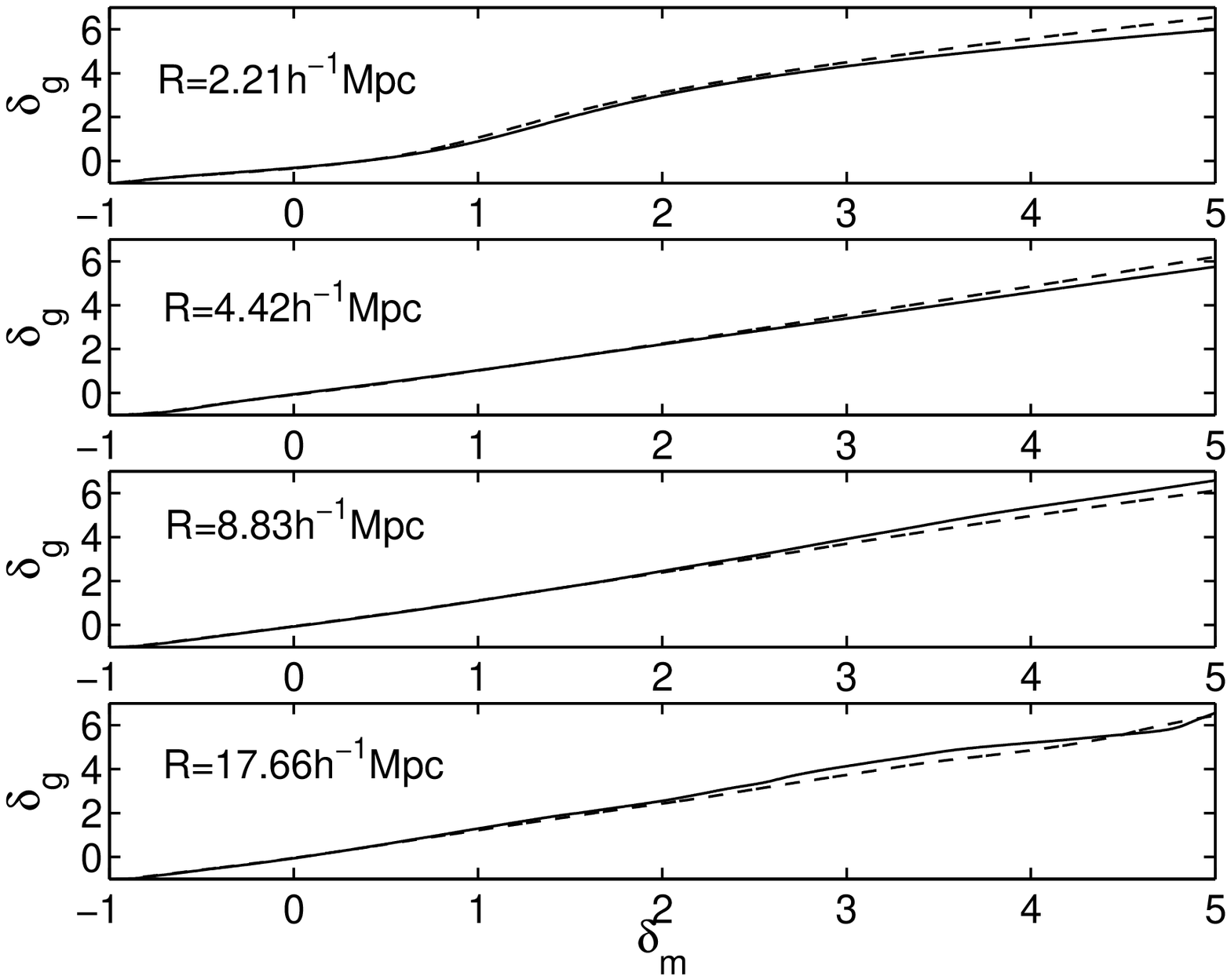}
\caption{Same as Figure 17, but for the mock galaxy catalog 
for the SDSS.}
\end{figure}
\section{Conclusion and Discussion}

We have presented a new method to extract the bias function
from galaxy catalogs. Since our method uses direct comparison
statistics extracted from simulations and data, 
it is applicable to a large range
of scales. In particular, its domain of validity includes the nonlinear
regime, which has proved to be impenetrable to other previous methods.
This is all the more important, since the most reliable data
are still available on non-linear scales. In addition, 
most available data pushing to the largest scales have still
significant non-linear ``contamination'', to which our method
is completely insensitive. In addition to expanding the range
of applicability, our method has an accuracy which rivals all
other methods. This is due to the fact, that it is using the
full counts in cells distribution, which is sensitive to 
very high order statistics.

The new technique is based on comparing the cumulative probability
distribution functions in simulations and data. To turn this
idea into a robust and reliable method, a difficult technical
challenge had to be met: the reconstruction of the continuous
probability distribution of density fluctuations from counts in 
cells measurements. This is a delicate, and potentially unstable
inversion, for which we have proposed two solutions. One is a
model independent inversion using a Richardson-Lucy iteration,
while the other is model dependent fit based
on the skewed lognormal approximation
(SLN3). The former method is useful down to scales where 
$\langle N \rangle\gsim 0.1$, and for relatively smaller
number of particles due to computational constrains.
The SLN3 fitting is useful possibly to
even slightly smaller scales, and it is feasible for large
simulations of arbitrarily high particle number.
Therefore the former lends itself naturally for fitting
the CPDF in galaxy catalogs, while the latter for large simulations.
The range of reconstruction of the bias function is $-1 \le \delta \le 5$,
and typically we could recover the bias from fairly realistic
simulations at the $5\%$ level. This suggests that our appication
of our method to contemporary catalogs, such as the SDSS and 2dF,
will constrain bias at an accuracy 
close to the absolute limit determined by systematic errors.

Most of our efforts have been centered on
reconstructing the bias function itself represents the mean bias.
We have found, however, that part of the
scatter is at least, perhaps dominantly, due to Poisson scatter.
Our reconstruction method corrects for this source of
error to the fullest possible extent. In addition, 
we have found that a simple shot noise model gives
excellent approximation to the stochastic component of the
bias.

The expected number of galaxies in region
with galaxy density fluctuation is $\delta_g$ is simply 
$N_g =\langle N_g \rangle (1+\delta_g)$. If we assume
a Poisson variance around this value (and neglect discreteness
in the dark matter catalog which is a good approximation)
we have
\begin{equation}
\frac{\Delta N_g}
{N_g }=
\frac{\Delta \delta_g}{1+\delta_g}=
\frac{1}{\sqrt{\langle N_g \rangle (1+\delta_g)}} \ . 
\end{equation}
It follows that
$\Delta \delta_g=\sqrt{(1+\delta_g)/\langle N_g \rangle}$.
This simple formula is shown in Figure 9, 11, 15 and 16
as a shaded area and provides an approximation to the
errorbars at typically at the 10\% level, with the largest
deviation being 50\%, mainly at large $\delta$. 
It appears that Poisson scatter
provides the dominant fraction of the stochasticity of
the bias. Note that there are signs of sub-Poisson
scatter in some of the Figures. The above is hardly
more than a toy model, and its degree of success 
is remarkable. 

A convenient parametrization of the (mean) bias function 
relies on a Taylor series expansion, 
\begin{equation}
\delta_g=f(\delta_m)=\sum_{k=0}^{\infty} \frac{b_k}{k!}\delta_m^k\ .
\end{equation} 
We adopted this form to fit our results empirically from the GIF and GIF-SDSS
mock catalogs for $\delta_m\in[-1,5]$. 
We have found that an expansion 
up to $k=2$ is always sufficient. The results, based on RL 
inversion, are shown in Table 4 \& 5.
It also quantifies the difference between redshift and real space.
For instance, $b=b_1$ is typically within a few percent
for the two cases.

A subtlety with the above formula is worth emphasizing again:
in this paper we have related the smoothed galaxy density
field to the smoothed dark matter density field. Other methods
might use ``unsmoothed'' field,  which really means that smoothing
is done on a much smaller scales than those scales considered
in the measurement. The meaning of a truly unsmoothed galaxy
density field, let alone its Taylor series expansion in terms 
of a truly unsmoothed dark matter density field is somewhat 
dubious. As well known, bias and smoothing does not commute,
which means that our bias coefficients might have slightly
different meaning than the quantities notated with the same
letters in other works.

\begin{deluxetable}{crrrrrrrrr}
\tablecolumns{8}
\tablecaption{Bias parameters for the GIF galaxy catalog}
\tablewidth{0pc}
\tablehead{
\colhead{} & \multicolumn{3}{c}{In Real Space} &\colhead{} &
\multicolumn{3}{c}{In Redshift Space} \\
\cline{2-4} \cline{6-8}\\
\colhead{ R($h^{-1}$Mpc)} &
\colhead{ $ b_0 $ } &
\colhead{ $ b_1 $ } &
\colhead{ $ b_2/2! $ } & 
\colhead{} &
\colhead{ $ b_0 $ } &
\colhead{ $ b_1 $ } &
\colhead{ $ b_2/2! $ } 
}
 \startdata
4.42  & 0.17 & 1.71  & -0.085 & \nodata &
       0.0029& 1.15   & 0.016 & \\
8.83  & 0.16 & 1.18  & -0.081 &  \nodata&
      0.10   & 1.17  & -0.066  \\
17.66 & 0.025 & 0.96  & -0.062 & \nodata&
        0.022 & 1.00  &  -0.047 \\
\enddata
\end{deluxetable}

\begin{deluxetable}{crrrrrrrrr}
\tablecolumns{8}
\tablecaption{Bias parameters for the mock SDSS galaxy catalog}
\tablewidth{0pc}
\tablehead{
\colhead{} & \multicolumn{3}{c}{In Real Space} &\colhead{} &
\multicolumn{3}{c}{In Redshift Space} \\
\cline{2-4} \cline{6-8}\\
\colhead{ R($h^{-1}$Mpc)} &
\colhead{ $ b_0 $ } &
\colhead{ $ b_1 $ } &
\colhead{ $ b_2/2! $ } &
\colhead{} &
\colhead{ $ b_0 $ } &
\colhead{ $ b_1 $ } &
\colhead{ $ b_2/2! $ } 
}
 \startdata
2.21  & -0.14 & 1.55 & -0.052 & \nodata &
      -0.11   & 1.57 & -0.037  \\
4.42  & -0.071 & 1.12 & 0.01  &  \nodata &
        -0.084 & 1.10 &  0.032  \\
8.83  & -0.062 & 1.20  & 0.032 &\nodata&
        -0.038   & 1.19  & 0.013 \\
17.66 & -0.015 & 1.20  & 0.059 & \nodata&
        0.005 & 1.19  &  0.014  \\
\enddata
\end{deluxetable}

No statistical method would be complete without a way of placing 
errorbars on the estimates derived from the method. Since our
method is fully non-linear and it relies on direct comparison
of galaxy catalogs with simulations, the only robust way of
producing errorbars is a Monte Carlo estimate. The procedure
is obvious: 
one has to repeat all the calculations
using a suit of simulations representing the data. As matter
of fact, we have demonstrated this in section 3.2, when we
have measured bias in several realizations of the randomly
diluted samples.  If the
dark matter simulation is large and dense enough, most of
the variance will come from data. While for most of our
present investigation only one realization was at our 
disposal, the CPU budget would allow the analysis of
a large number of simulations.
It took about one hour to measure CIC in
dark matter and galaxy catalogs, about 30 minutes for one SLN3 fit,
and up to a few hours for RL inversion (typical value in our calculation
of the mock SDSS). We have used a 2.4GHz dual Xeon workstation with
2GBytes of  memory.

In this work we exclusively used the Poisson model to relate
the discrete galaxy distribution to the underlying continuous
field. This approximation might break down, especially on
very small scales, where halo models are known to predict
sub-Poisson scatter \citep{cmsb02,bw02}. Such theories, once firmly
established, can be naturally incorporated into our formalism
via simply generalizing the kernel. Another application,
where the kernel will need modification is recovery
of the bias from two dimensional angular catalogs. In that
case the corresponding kernel would depend on the selection
function, but the core idea of the method would still work.
These generalizations as well as applications 
to real data are in preparation.

\acknowledgments

This work was supported by NASA through grants AISR NAG5-11996,
ATP NAG5-12101, and by NSF through grants AST02-06243 and 
ITR 1120201-128440.
The simulations in this paper were carried out by the Virgo
Supercomputing Consortium using computers based at Computing Centre of 
the Max-Planck Society in Garching and at the Edinburgh Parallel 
Computing Centre. The data are publicly available at 
{\bf http://www.mpa-garching.mpg.de/Virgo}.

\end{document}